\documentclass[aps,preprint,amsmath,amssymb,floatfix]{revtex4-1}
\usepackage{graphicx,morefloats,float}
\usepackage{color}
\usepackage{amsmath}
\usepackage{bbm}
\def\Re{\text{Re}}
\def\Im{\text{Im}}
\def\sgn{\text{sgn}}
\newcommand{\KFS}{\text{K$_{1-x}$Fe$_{2-y}$Se$_2$}}
\newcommand{\KTFS}{\text{(K,Tl)$_{1-x}$Fe$_{2-y}$Se$_2$}}
\newcommand{\TFS}{\text{Tl$_{1-x}$Fe$_{2-y}$Se$_2$}}

\begin{document}
\hyphenation{va-ni-sh-ing}

\begin{center}
{\large\bf Superconductivity at the Border of Electron Localization and Itinerancy}
\\[0.5cm]

Rong Yu$^{1,2\ast}$, Pallab Goswami$^{2,3\ast}$, Qimiao Si$^{2}$, Predrag Nikolic$^4$,
and Jian-Xin Zhu$^5$ \\

$^1$ Department of Physics, Renmin University of China, Beijing 100872, China

$^2$ Department of Physics and Astronomy, Rice University, Houston,
TX 77005, USA\\

$^3$ National High Magnetic Field Laboratory, Florida State University, Tallahassee, Florida 32310, USA\\

$^4$
 School of Physics, Astronomy and Computational Sciences,
 George Mason University, Fairfax, VA 22030, USA\\

$^5$ Theoretical Division and Center for Integrated Nanotechnologies,
Los Alamos National Laboratory, Los Alamos, NM 87545, USA\\

$^{\ast}$ These authors contributed equally to this work

\end{center}

\vspace{0.5cm}
{\bf
The superconducting state of  iron pnictides and chalcogenides exists at the border of antiferromagnetic order.
Consequently, these materials could provide clues about the relationship between magnetism
and unconventional superconductivity. One explanation, motivated by the so-called bad-metal behaviour
of these materials, proposes that magnetism and superconductivity develop out of quasi-localized
magnetic moments which are generated by strong electron-electron correlations. Another suggests that these phenomena
are the result of weakly interacting electron states that lie on nested Fermi surfaces.
Here we address the issue by comparing the newly discovered alkaline iron selenide superconductors,
which exhibit no Fermi-surface nesting, to their iron pnictide counterparts. We show that the strong-coupling
approach leads to similar pairing amplitudes in these materials, despite their different Fermi surfaces.
We also find that the pairing amplitudes are largest at the boundary between
electronic localization and itinerancy, suggesting that new superconductors might be found in materials
with similar characteristics.
}

\newpage

Superconductivity often occurs
near
a magnetic order. This is the case not only in the iron
based compounds \cite{Kamihara08,PCDai,MQazilbash},
but also in heavy fermion intermetallics, organic charge-transfer salts and copper  oxides.
An important question that is central to these diverse classes of
unconventional superconductors
is whether the mechanism of their superconductivity is in a way analogous to that of conventional
superconductors,
with spin fluctuations replacing phonons as
the glue for electron pairing,
or
it instead involves novel electronic states that are generated by strong electron
correlations \cite{Anderson}.

Iron-based superconductors
represent a unique setting to elucidate this basic issue:
their large materials parameter space
provides the opportunity to understand the microscopic physics
by comparing the properties across their
material families.  A major recent development suitable for this important characteristics
is the discovery of high temperature
superconductivity in a new family of iron chalcogenides, the alkaline iron selenides \KFS\
(Ref.~\onlinecite{JGuo}). Other related iron selenides, with $\mathrm{K}$ replaced in part or in entirety by
$\mathrm{Rb}$, $\mathrm{Cs}$ or  $\mathrm{Tl}$, behave similarly \cite{MFang,LSun}.
The key property is that the maximum of the superconducting transition temperature ($T_c$) observed
in the alkaline iron selenides, above 30 K, is similar to that
of their 122 iron pnictides counterpart, suggesting a commonality in the underlying mechanism
for superconductivity across these systems.

Compared to those of the iron pnictides, the Fermi surfaces  in the alkaline iron selenides
are very different.
While the former contain both electron and hole pockets, respectively at the boundary
(M) and center ($\mathrm{\Gamma}$) of the one-Fe Brillouin zone, only electron pockets
are present in the alkaline iron selenides \cite{YZhang, TQian, DMou}.
The weak-coupling Fermi-surface-based mechanism will operate very differently
in the alkaline iron selenides compared to the iron pnictides
\cite{TAMaier, FWang, MKhodas}.

Here we demonstrate that,
by incorporating the bad-metal nature of the normal state,
the strong-coupling approach provides the
understanding
for the comparable pairing strength in the
alkaline iron selenides
and iron pnictides.

\noindent
{\bf Results}
\\
{\bf Proximity to Mott Transition.}
We seek for the commonality between the iron chalcogenides and pnictides
based on the observation that the parent compound
of the alkaline iron selenides are antiferromagnetic insulators \cite{MFang,DMWang}.
These insulating  selenides contain Fe-vacancies that are ordered in the Fe-square lattice,
so that the Fe valence is kept at $+2$. Because of their very large ordered moment
of  about 3.3 $\mu_B$/Fe (Refs.~
\cite{WBao1,MWang}), they are naturally considered
as Mott insulators, arising through a kinetic-energy reduction induced by the ordered
vacancies \cite{RYu1,YZhou}. An important question is how $U/W$, the ratio of a combined
onsite Coulomb and Hund's interaction to the electron bandwidth, compares with the Mott-transition
threshold, $U_{\text{c}}/W$. Given that a modest reduction of the kinetic energy from the parent
iron pnictides to the parent alkaline iron selenides turns the system from metallic to insulating,
we can infer that $U/W$ is larger than but close to  $U_{\text{c}}/W$  in the alkaline iron selenides,
while smaller than but also close to $U_{\text{c}}/W$ in the iron pnictides. Hence, both superconductors
arise out of bad metals on the verge of a Mott localization.

This is illustrated in Fig.~1, showing the parent compounds
of the alkaline iron selenides and iron pnictides
in the vicinity of, {\it albeit} on the two sides of,  the Mott transition point.
Consequently, we use the Mott transition point as anchoring the regime of the phase diagram
that has strong antiferromagnetic correlations, illustrated by the purple shading in Fig.~1.
At the Mott transition point, all the electronic excitations are incoherent. Integrating out
the gapped electronic excitations gives rise to
a model of localized spins with nearest-neighbor ($J_1$) and next-nearest-neighbor
($J_2$) interactions on the Fe-square lattice.
In the carrier doped regime, a five-band $t$-$J_1$-$J_2$
model ensues~\cite{QSi1,QSi2,PGoswami}.
More generally, for the iron pnictides, the proximity to the Mott transition has
also been supported by the bad-metal behavior of the normal state, as determined by
the optical conductivity \cite{MQazilbash} and other measurements, providing  the basis
for strong-coupling
approaches to superconductivity
\cite{QSi1, QSi2, Haule08, KSeo,PGoswami,Moreo,Berg,Lv}.

\noindent
{\bf Multi-orbital $t$-$J_1$-$J_2$ Approach.}
We study spin singlet pairings in
two such five-band $t$-$J_1$-$J_2$ models,
respectively for the alkaline iron selenides and iron pnictides.
The different Fermi surfaces arise from different choices of the tight-binding parameters,
which are specified by the kinetic energy part of the model.
The five $3d$ orbitals of iron
are
used in order to correctly describe the Fermi surfaces;
they are denoted by $\alpha=1,...,5$,
which correspond to
$3d_{xz}$, $3d_{yz}$, $3d_{x^2-y^2}$,
$3d_{xy}$, and $3d_{3z^2-r^2}$
(see Methods).
The Fermi surface of
\KFS\ is shown in Fig.~2a. It comprises electron pockets only, and corresponds to
an electron doping of about $15\%$ per Fe; both have been specified in accordance with
the angle-resolved photoemission spectroscopy (ARPES) measurements \cite{YZhang, TQian, DMou}.
The electron and hole like Fermi pockets
for the iron pnictides \cite{Graser}, also with $\delta =15\%$ electron doping,
are displayed in Fig.~2b.
 The $xz/yz$ and $xy$ $3d$
orbitals dominate the electronic states near the Fermi surfaces, as illustrated in Figs.~2c,d.
We observe that,
at and near zero doping ($\delta=0$), the ground state will be antiferromagnetically
ordered (see below). We also note that our study focuses on the couplings in the spin sector
as driving the superconductivity, although our general analysis may also have implications
for the considerations in the orbital sector \cite{Lv}.

Figures~3a,b demonstrate how superconductivity in the five-band
$t$-$J_1$-$J_2$
is magnetically driven by the short-range $J_1$-$J_2$ exchange interactions.
Shown here, respectively for
\KFS\ and iron pnictides, are the zero-temperature phase diagrams
in the $J_1$-$J_2$ plane, for $0 \le J_1$ and $J_2 \le 0.3D$,
where $D$ is the renormalized bandwidth.
For \KFS\,  the $J_2$ dominated region I has
$A_{1g}$ symmetry, with $s_{x^2y^2}$ ($\cos k_x \cos k_y$) wave
being the dominant pairing component.
Regions II, III, and IV are primarily of the
$B_{1g}$ $d_{x^2-y^2}$ $(\cos k_x- \cos k_y)$ wave character.
The distinction among the three regions reflects the difference in the admixture of a small
$A_{1g}$ component at zero (and low) temperatures
(see Methods).
This is similar to the phase diagram of the pnictides case, where the
$J_2$ dominated region I also is primarily $A_{1g}$ $s_{x^2y^2}$ wave,
and regions II and III are primarily
$B_{1g}$ $d_{x^2-y^2}$ wave.

To clarify the admixture of the different pairing components, we show the evolution
of the amplitudes of these components,
projected onto the $3d_{xy}$ orbital, in Figs.~3c and 3d.
Comparing the two figures clearly shows that, for the alkaline iron selenides,
the dominant pairing amplitudes in the  $s_{x^2y^2}$  and $d_{x^2-y^2}$
channels are
comparable to their counterparts in the iron pnictides. This is further illustrated in Figs.~4a,b,
which show the pairing amplitudes, also projected to the $3d_{xy}$ orbital,
vs. $J_1 / D$ for a fixed $J_2 / D =0.1$.
The same conclusion applies to the pairing amplitudes projected to the other 3d orbitals,
as shown in Figs.~4c,d
for the case of $3d_{xz}/3d_{yz}$ orbitals.

{\bf Pairing Amplitudes in Alkaline Iron Selenides.}
We therefore reach the conclusion that the pairing amplitudes
in models
respectively
for the alkaline iron selenides
and iron pnictides are similar for given dimensionless exchange interactions,
$J_1 / D$ and $J_2 / D$.
This is surprising, because the alkaline iron selenides lack
any hole Fermi pocket and, therefore, do not possess any Fermi-surface nesting.
Our result is inherent to the strong coupling approach; here, while the details of the
Fermi surfaces are important,
the superconducting pairing does not require coupling between hole- and electron-
Fermi pockets.
Instead, the driving force for the pairing lies in the close-neighbor exchange interactions,
$J_1$ and $J_2$. The presence of electron pockets near
the M points of the Brillouin zone is adequate to promote the
$\cos k_x \cos k_y$ $A_{1g}$ $s_{x^2y^2}$-wave pairing,
as well as the $\cos k_x - \cos k_y$ $B_{1g}$ $d_{x^2-y^2}$-wave pairing.
For similar ratios of
$J_1/D$ and $J_2/D$, the corresponding
pairing amplitude
is naturally comparable to that of the iron pnictides,
in which both the electron pockets near
M and the hole pockets near
$\Gamma$ promote these two pairing components.

{\bf Enhancement of Pairing Amplitudes near Mott transition.}
The results for both classes of materials (Figs.~3c,d and 4) also show that
the pairing amplitudes are larger when
$J_1/D$ and $J_2/D$ are increased. This conclusion, in turn, leads to a general principle.
To see this, we note that the exchange interactions increase as the Mott transition is
approached from the insulating side ({\it cf.}, Fig.~1), because of the reduction
of the charge gap. At the same time, the renormalized bandwidth $D$ is reduced as the
Mott transition is approached from the metallic side. Correspondingly, at the boundary
between electronic localization and delocalization,
the ratios $J_1/D$ and $J_2/D$ will be maximized and so will the superconducting
pairing amplitudes.

Our results  provide the understanding for some key experimental observations.
Inelastic neutron scattering experiments have shown that the exchange interactions
in the alkaline iron selenides and iron pnictides have the same order of magnitude~\cite{MWang}.
Furthermore, as Fig.~1 illustrates, the alkaline iron selenides and iron pnictides have
approximately the same doping concentration and similar degree of proximity to the Mott transition.
Therefore, $J_1/D$ and $J_2/D$ are similar in the two materials.
This leads to our key conclusion, namely the two classes of iron based superconductors
have comparable pairing amplitudes and, by extension, comparable superconducting
transition temperatures.

It is instructive to contrast the situation here with
$\mathrm{K}\mathrm{Fe}_2\mathrm{As}_2$. The latter system, with the Fermi surface
containing only hole pockets~\cite{Sato09},  represents another material lacking
Fermi-surface nesting. However, it is strongly doped,
with a hole doping of 0.5 per Fe.
This extreme overdoping in $\mathrm{K}\mathrm{Fe}_2\mathrm{As}_2$
means that the system is far away from the Mott transition anchoring point
discussed here and should, therefore, have considerably reduced pairing amplitudes.
Experimentally, it indeed has a much smaller $T_{\text c}$  of about 3 K.

For the alkaline iron selenides, our results show dominating
$A_{1g}$ $s_{x^2y^2}$ and $B_{1g}$ $d_{x^2-y^2}$-wave states in competition.
Both have nearly isotropic and nodeless gaps on the electron pockets, and this is consistent
with all existing measurements
of the superconducting gap.
Neutron scattering experiments have identified a resonance associated with the superconducting
state~\cite{Park12}.  This is most readily understood in terms of
a $B_{1g}$ $d_{x^2-y^2}$-wave pairing~\cite{Park12},
although it may also arise from an $A_{1g}$ $s_{x^2y^2}$ channel
once the effect of two iron ions per unit cell
is taken into account~\cite{Xu12}. On the other hand, ARPES experiments
have suggested that the superconducting gap is nodeless on the faint electron pocket near the $\Gamma$ point~\cite{Xu12,Wang12}.
This appears to favor an $A_{1g}$ $s_{x^2y^2}$-wave pairing~\cite{Xu12,Wang12}.
Further experiments are needed to settle which of the two possible pairing channels
operates in the alkaline iron selenides.
For the iron pnictides, the dominance of $A_{1g}$ $s_{x^2y^2}$ state
in a large portion of the phase diagram
is consistent with the sign-changing $s_\pm$
paired state arising in both weak and strong coupling approaches.
Compared to the alkaline iron selenides, an important difference is that
the pairing amplitude in the $A_{1g}$ $s_{x^2+y^2}$ channel is sizable
(Figs.~3,4).
This arises because the corresponding form factor, $(\cos k_x + \cos k_y)$, while negligible
at the dominant electron Fermi surfaces located near the M points for the alkaline iron selenides,
is large near the hole Fermi pockets around the $\Gamma$ point in the case of the iron pnictides.
The relevance of the $A_{1g}$ $s_{x^2+y^2}$ channel
 is important for understanding a possible strong momentum dependence or even the development
 of nodes in the superconducting gap;
nodes arise on the electron Fermi pockets near the M points
when the amplitude of the $A_{1g}$ $s_{x^2+y^2}$ component exceeds
a threshold value
compared to that of the coexisting $A_{1g}$ $s_{x^2y^2}$ component.
Indeed, experimentally, some iron pnictide superconductors show fully-gapped behavior,
while others display properties that suggest the existence of nodes.

\noindent
{\bf Discussion}
\\
The bad metal behavior near the Mott transition
also pertains to the relationship between magnetism and
superconductivity
in the alkaline iron selenides. For the vacancy-ordered
parent insulating system (the so-called 245 phase),
it causes the electronic excitations to have a large incoherent component;
the latter gives rise to
a large spin spectral weight even in the absence of any itinerant carriers, as have been
 observed by neutron scattering experiments \cite{WBao1,MWang}.

In superconducting compounds, there is direct evidence for a phase separation \cite{NATPHYS}
between a superconducting component and the parent insulating antiferromagnetic part.
The superconducting component is free of ordered vacancies \cite{NATPHYS},
suggesting a tetragonal 122 structure as we have used in our model.
Our study here focuses on the pairing instabilities in the underlying
metallic state of this component,
while taking advantage of its connection
with the Mott insulating
phase  in the overall phase diagram \cite{YuSi_OSMT}
discussed in the Supplementary Note 3  (and Supplementary Figure S5)
and evidenced by recent experiments \cite{Yi_OSMT,Gao_OSMT,WLi_OSMT}.

It is worth emphasizing that
the physical pathway
linking the superconducting phase and 245 Mott-insulating phase involves varying both
vacancy order and carrier concentration.
Indeed,
the existence of the multiple phases of the alkaline iron selenides
suggests  an overall, extended, parameter space
in which the different phases can be connected.
This is described in some detail in the Supplementary Note 3 and is in particular illustrated by
 Supplementary Figure S5.
 The details of this physical pathway is
a distinct issue that is intriguing and important
in its own right
\cite{Yi_OSMT,YuSi_OSMT},
and other alkaline iron selenides such as
KFe$_{\rm 1.5}$Se$_{\rm 2}$,
with one vacancy per four iron sites \cite{Zhao12,MFang}
(and small carrier doping \cite{Yi_OSMT,Chen11}),
and  ${\rm K_{0.5}Fe_{1.75}Se_2}$, with
one vacancy per eight iron sites  \cite{Wen13,CaoZhang},
may also be placed  in this extended phase diagram.
For the purpose of the present work, what is particularly pertinent
is the clue
these multiple phases have provided for the strength of the
electron correlations. As described in the introduction and illustrated in Fig.~1,
the existence of the Mott-insulating phase
in the 245 compound suggests not only
that the underlying
$U/W$ is larger than $U_{\text c}/W$ in the vacancy-ordered alkaline iron selenides,
but also that $U/W$ is below but close to $U_{\text c}/W$ in the iron pnictides.
This proximity to the Mott transition allows for the present study on the
pairing amplitudes in both classes of iron-based superconductors.

Recently, superconductivity of around 50 K was reported in a single-layer FeSe film grown
on SrTiO$_3$ substrate~\cite{FeSeSTO}. ARPES measurements~\cite{DLiu12} indicate
that the Fermi surface consists of only electron pockets, similar to that of the alkaline
iron selenides. Thus, it is instructive to model the single-layer FeSe film and compare
its pairing properties with those of the other iron-based superconductors.
We have thus studied the singlet superconductivity in
 a similar five-orbital $t$-$J_1$-$J_2$ model for the single-layer FeSe film (see Methods),
 with an electron doping of $0.1$ per Fe;
 we consider the role of the substrate as providing the structure which dopes
 this amount of electron carriers into the single-layer FeSe
 ~\cite{STan13,SHe13}.
The results are shown in Supplementary Figures S7 and S4.d.
The pairing phase
diagram of the single-layer FeSe is comparable to those for both the alkaline iron selenides
and iron pnictides, and so are the overall pairing amplitudes.

By showing that pairing amplitudes are similar
for the iron chalcogenides and pnictides
in spite of their drastically different Fermi surfaces,
our results
uncover a universality in the existing and emerging
iron-based high temperature
superconductors
with very diverse materials and Fermi-surface characteristics. Moreover, our demonstration, that
that the pairing amplitudes increase with the ratio of the short-range spin exchange energy
to the renormalized kinetic energy,
reveals an important principle.
Namely, superconductivity is optimized at the border between
itinerancy and electronic localization. This principle
should apply
beyond the context of iron pnictides and chalcogenides,
and is expected to guide the search for
superconductors
with even higher transition temperatures.

\vskip 0.2 cm

\noindent
{\bf Methods}
\\


\noindent{\bf
Multi-orbital $t$-$J_1$-$J_2$ Model}

We describe the methods used in our study of the phase diagram for
the singlet superconducting pairing of five orbital
$t-J_1-J_2$ models,
described as $H=H_t+H_J$. Here, $H_t$, and $H_J$ respectively correspond
to the fermion hopping
and the $J_1-J_2$ exchange Hamiltonians.

Two such models are considered.
The difference in the fermiology of the alkaline iron selenides and iron pnictides are specified
via the kinetic
part $H_t$, and we have chosen the tight binding parameters by fitting the band dispersions obtained
from density functional calculations.
The short-range antiferromagnetic exchange interactions in $H_J$
drive the singlet pairings.
We have chosen the same exchange coupling constants $J_1$, $J_2$ for each orbital, and performed
a mean field decoupling
of the exchange part in the pairing channel.
Each orbital contributes four pairing amplitudes $\Delta_{a,\alpha}$,
which are respectively given by $s_{x^2+y^2,\alpha}$, $d_{x^2-y^2,\alpha}$, $s_{x^2y^2,\alpha}$,
$d_{xy,\alpha}$,
where $\alpha=1,...,5$ labels the orbitals; these amplitudes are related to their real-space counterparts,
$\Delta_{{\mathbf{e}},\alpha}
=\langle c_{i\alpha\uparrow}c_{i+{\mathbf{e}}\alpha \downarrow}
-c_{i\alpha\downarrow}c_{i+{\mathbf{e}}\alpha \uparrow}\rangle/2$.
We have minimized the free energy to find the self-consistent solution for the twenty pairing
amplitudes.
We now expound on these in some detail.

Our
five-orbital $\mathrm{t}$-$\mathrm{J}_1$-$\mathrm{J}_2$ model
arises from an expansion of the five-orbital Hubbard model with respect to the
Mott transition point (the ``$w$-expansion") \cite{QSi2}.
The Hamiltonian for the model
is given by
\begin{eqnarray}
H&=&-\sum_{i<j,\alpha,\beta,s} t_{ij}^{\alpha \beta}
c^{\dagger}_{i\alpha s}c_{j\beta s}+h.c.
-\mu \sum_{i,\alpha}n_{i\alpha} \nonumber \\
&&+\sum_{\langle ij\rangle,\alpha,\beta} J_{1}^{\alpha \beta}
\left(\vec{S}_{i\alpha}\cdot \vec{S}_{j\beta}-\frac{1}{4}n_{i\alpha} n_{j\beta}\right)
+\sum_{\langle \langle ij\rangle \rangle,\alpha,\beta} J_{2}^{\alpha \beta}
\left(\vec{S}_{i\alpha}\cdot \vec{S}_{j\beta}-\frac{1}{4}n_{i\alpha} n_{j\beta}\right),
\end{eqnarray}
where $c^{\dagger}_{i\alpha s}$
creates an electron at site $i$, with orbital index $\alpha$ and spin
projection $s$; $\mu$ is the chemical potential and
$t_{ij}^{\alpha \beta}$ the hopping matrix. The orbital index $\alpha=1,2,3,4,5$ respectively correspond to
five $3\mathrm{d}$ orbitals $3\mathrm{d}_{xz}$, $3\mathrm{d}_{yz}$, $3\mathrm{d}_{x^2-y^2}$,
$3\mathrm{d}_{xy}$, and $3\mathrm{d}_{3z^2-r^2}$ of iron.
The nearest-neighbor (n.n., $\langle ij\rangle$)
and next-nearest-neighbor (n.n.n., $\langle \langle ij \rangle \rangle$)
exchange interactions are respectively denoted by $J_{1}^{\alpha \beta}$ and $J_{2}^{\alpha \beta}$.
The spin operator
$\vec{S}_{i\alpha}=\frac{1}{2}\sum_{s,s^{'}}c^{\dagger}_{i\alpha s}
\vec{\sigma}_{ss^{'}}c_{i\alpha s^{'}}$
and the density operator $n_{i \alpha}=\sum_{s}c^{\dagger}_{i\alpha s}c_{i\alpha s}$,
with $\vec{\sigma}$ representing the
Pauli matrices.

For the calculation in these five-orbital models,
we have considered the effects of the fermion no double-occupancy
constraints as being implicitly
accounted for by the reduction of the effective bandwidth $D$.
We expect that this treatment
 does not  affect the results qualitatively.
We have verified this expectation for a two orbital model, by explicitly
keeping track of the occupancy constraints.
This is described this in some detail in the Supplementary Note 1, Supplementary Methods,
and Supplementary Figure S3.
As shown there,
the phase diagram and pairing amplitudes are insensitive to this when the exchange interactions are measured w.r.t.
the renormalized bandwidth $D$.

In a previous study of a multi-orbital $\mathrm{t}-\mathrm{J}_1-\mathrm{J}_2$  model
for iron pnictides \cite{PGoswami},
it has been demonstrated by three of the present authors that the dominant pairing symmetry
is governed by the intra-orbital
exchange interactions, and the consideration of the orbitally off-diagonal exchange interaction
only introduces quantitative
modifications of the phase diagram.
Therefore to keep the analysis simple, we will consider the orbitally diagonal exchange couplings
$J_{1}^{\alpha \beta}=\mathrm{J}_{1}\delta_{\alpha \beta}$ and
$J_{2}^{\alpha \beta}=\mathrm{J}_{2}\delta_{\alpha \beta}$.
To consider a pure superconducting phase, we will study here the case with the 122 tetragonal symmetry
(see
Supplementary Note 3 for further discussions).

To describe the fermiology, we use a tetragonal symmetry preserving tight binding model,
involving all five $3\mathrm{d}$ orbitals of iron. In the momentum space the $5\times5$ tight-binding matrix,
and its eigenvalues will be respectively denoted by $\hat{\xi}_k$, and $E_{j,\mathbf{k}}$.
The total number of electrons is determined by the chemical potential and the dispersion relations,
according to $ n=2\sum_{j, \mathbf {k}} \theta(\mu-E_{j,\mathbf{k}})$, and carrier doping $\delta=|n-6|$.
The tight-binding parameters are fixed by fitting the band structure obtained from the LDA calculation.
The details of the tight-binding parametrization for iron chalcogenides and the associated
band dispersions are discussed
below, in
the next subsection.
For
\KTFS\
the band dispersions correctly produce the electron pockets near the
$M$ points, as illustrated in Fig.~2a for
\text{K$_{1-x}$Fe$_{2-y}$Se$_2$}.
The electron and
hole like Fermi pockets obtained  from a similar five orbital tight-binding model of iron pnictides \cite{Graser}
are shown in Fig.~2b,
to contrast the fermiology of the two materials.
The fermiology of
\TFS\ is shown below, in Supplementary Note 2 and Supplementary Figures S1 and S2,
which again consists of only electron pockets. We note that
ARPES experiments
\cite{YZhang, TQian, DMou}
have suggested that very weak electron-like pockets may also exist near the $\Gamma$ points.
Unlike their hole counterpart in the iron pnictides,
these electron pockets have very small spectral weight and are
expected to play at most a secondary
role in driving superconductivity.

\vspace{0.2cm}
\noindent{\bf Details of tight-binding parametrization}
\label{tight-binding}

For the five orbital, tetragonal symmetry preserving tight binding model,
we adopt the parametrization method of Ref.~\onlinecite{Graser}.
We have fitted the LDA
band structure with the band dispersion found from the tight-binding model, to determine the hopping parameters.
For
\text{(K,Tl)$_{1-x}$Fe$_{2-y}$Se$_2$},
we have performed calculations with two different sets of hopping parameters
(see Supplementary Tables S1 and S2).
These two sets respectively were derived from fitting the LDA results
for  $\mathrm{K}\mathrm{Fe}_2\mathrm{Se}_2$, and
$\mathrm{Tl}\mathrm{Fe}_2\mathrm{Se}_2$;
see Supplementary Figure~S1.
For iron pnictides we use the tight-binding parameters of Ref.~\onlinecite{Graser}.
In
Fig.~2a and 3a,
we have shown the electron pockets and pairing phase diagram for the
band structure corresponding to
\text{K$_{1-x}$Fe$_{2-y}$Se$_2$}.
In
Supplementary Figure~S1,
we provide the band dispersions for both tight-binding models,
and also show the electron pockets derived from
\TFS\ .

\vspace{0.2 cm}
\noindent{\bf Spin-singlet pairing states}

We first consider degenerate
pairing states in the absence of kinetic energy.
The spin singlet, intraorbital pairing operators are defined
as $\Delta_{{\mathbf{e}},\alpha} \equiv \Delta_{{\mathbf{e}},\alpha \alpha}
=\langle c_{i\alpha\uparrow}c_{i+{\mathbf{e}}\alpha \downarrow}
-c_{i\alpha\downarrow}c_{i+{\mathbf{e}}\alpha \uparrow}\rangle/2$,
where ${\mathbf{e}}=\hat{x},\hat{y},\hat{x}\pm\hat{y}$. Two types of pairing states $\Delta_{\hat{x},\alpha}=\pm \Delta_{\hat{y},\alpha}$,
respectively denoted as $s_{x^2+y^2}$ and $d_{x^2-y^2}$ states with momentum space form factors
$g_{x^2+y^2,\mathbf{k}}=\cos k_x + \cos k_y$, $g_{x^2-y^2,\mathbf{k}}=\cos k_x - \cos k_y$ arise
due to the nearest neighbor exchange interaction, and they are energetically degenerate
in the absence of kinetic energy.
Similarly the next nearest neighbor exchange gives rise to two degenerate pairing states
$\Delta_{\hat{x}+\hat{y},\alpha}=\pm \Delta_{\hat{x}-\hat{y},\alpha}$,
respectively denoted as $s_{x^2y^2}$ and $d_{xy}$ states,
with momentum space form factors $g_{x^2y^2,\mathbf{k}}=\cos (k_x -k_y)+ \cos (k_x+ k_y)$,
$g_{xy,\mathbf{k}}=\cos (k_x -k_y)+ \cos (k_x+ k_y)$. A strong magnetic frustration,
characterized by $J_1\sim J_2$,
leads to an enhanced degeneracy among the four paired states.

The kinetic energy term lifts most of these degeneracies.
In the strong frustration regime ($J_1\sim J_2$)
leaves a quasi-degeneracy among a reduced set of pairing states.
To study the full problem we perform a mean-field decoupling \cite{Kotliar,PGoswami}
 of the exchange interaction terms. We introduce four complex singlet pairing amplitudes for each orbital,
 and write the following $5 \times 5$ pairing matrix $\mathbf{\Delta}_{\mathbf{k}}=\sum_{a}\mathrm{diag}[\Delta^{a}_{{\mathbf{k}},11},
\Delta^{a}_{{\mathbf{k}},22},\Delta^{a}_{{\mathbf{k}},33},\Delta^{a}_{{\mathbf{k}},44},
\Delta^{a}_{{\mathbf{k}},55}]$, where $\Delta^{a}_{{\mathbf{k}},\alpha \alpha}=\Delta^{a}_{\alpha \alpha}g_{a,\mathbf{k}}$,
and the index $a$ corresponds to $s_{x^2+y^2}$, $d_{x^2-y^2}$, $s_{x^2y^2}$
and $d_{xy}$ symmetries. In the subspace of $xz$ and $yz$ orbitals, which transform as a doublet under the tetragonal point group operations,
there are the following four classes of intra-orbital pairing states for an orbitally
diagonal
$\mathrm{J}_1$-$\mathrm{J}_2$
 model:
(i) $\mathrm{A_{1g}}:[s_{x^2+y^2}^{A_{1g}}g_{x^2+y^2,\mathbf{k}}
+s_{x^2y^2}^{A_{1g}}g_{x^2y^2,\mathbf{k}}]\tau_0
+d_{x^2-y^2}^{A_{1g}}g_{x^2-y^2,\mathbf{k}}\tau_z$;
(ii) $\mathrm{B_{1g}}:d_{x^2-y^2}^{B_{1g}}
g_{x^2-y^2,\mathbf{k}}\tau_0
+[s_{x^2+y^2}^{B_{1g}}g_{x^2+y^2,\mathbf{k}}
+s_{x^2y^2}^{B_{1g}}g_{x^2y^2,\mathbf{k}}]\tau_z$;
(iii) $\mathrm{A_{2g}}: d_{xy}^{A_{2g}}g_{xy,\mathbf{k}}\tau_z$;
and (iv) $\mathrm{B_{2g}}: d_{xy}^{B_{2g}}g_{xy,\mathbf{k}}\tau_0$. The eight pairing amplitudes
$s_{x^2+y^2}^{A_{1g}}$ etc. are obtained as linear combinations of $\Delta^{a}_{11}$
and $\Delta^{a}_{22}$, which are in turn linear combinations of $\Delta_{{\mathbf{e}},11}$,
and $\Delta_{{\mathbf{e}},22}$.

\vspace{0.2cm}

\noindent{\bf Mixed-symmetry pairing states
breaking time-reversal symmetry}

The admixed states II and III
of Figs.~3a,b
break time reversal symmetry and have the form $A_{1g}+iB_{1g}$. In regions II and III,
the $A_{1g}$ components are respectively $\mathrm{s}_{x^2y^2}$ $\cos k_x \cos k_y$
and $\mathrm{s}_{x^2+y^2}$ $(\cos k_x +\cos k_y)$.
In contrast to the pnictides, the pairing phase diagram of 122 chalcogenides has a region IV,
which is of purely $B_{1g}$ $\mathrm{d}_{x^2-y^2}$ character, even at zero temperature.
The time-reversal symmetry breaking in region II is due to quasi-degeneracy between
$s_{x^2y^2}$ and $d_{x^2-y^2}$ pairings,
induced by strong magnetic frustration. On the other hand, the quasi-degeneracy between
$s_{x^2+y^2}$ and $d_{x^2-y^2}$
pairings in region III arises due to band-with suppression. These mixed-symmetry pairing states
are expected to be relevant only at sufficiently low temperatures.

\vspace{0.2cm}

\noindent{\bf Generalizations of the models}

The pairing phase diagrams,  shown in
Figs.~3a,b,
have been determined by assuming orbitally
diagonal exchange couplings $J_1$ and $J_2$. The inter-orbital exchange couplings do not qualitatively
modify the phase diagram.  The inter-orbital couplings
introduce some sub-dominant components to $A_{1g}$ and $B_{1g}$ regions,
while leaving the competition
between $A_{1g}$ and $B_{1g}$ pairing symmetries intact.
For example an inter-orbital second neighbor coupling
between $xz$ and $yz$ orbitals gives rise to inter-orbital $d_{xy} (\sin k_x \sin k_y)$ component,
which is a part of $A_{1g}$ pairing. Similarly we can also consider further neighbor
intra-orbital exchange couplings.
For example the third neighbor antiferromagnetic exchange coupling $J_3$ does
not change the competing pairing channels;
but introduces sub-dominant $A_{1g}$ $(\cos 2 k_x + \cos 2k_y)$ and $B_{1g}$ $(\cos 2k_x -\cos 2k_y)$ components.
Therefore the the phase diagram obtained from an orbitally diagonal $J_1-J_2$ model is generic
and robust.

We also note that an extended $J_1-J_2$ model \cite{YuGoswami11,MWang},
with ferromagnetic
nearest neighbor coupling,
defined on a modulated square lattice, has been used to explain the $\sqrt{5}\times \sqrt{5}$ block
spin antiferromagnetic order. While this is believed to reflect the modulation to the exchange
interactions that exists only
in the presence of ordered vacancies \cite{YuGoswami11,CaoDai11},
it is interesting to consider the effect of a ferromagnetic $J_1$.
The latter will suppress the $B_{1g}$ pairing, and increase the $A_{1g}$ region.
We stress that the $A_{1g}$-$B_{1g}$
competition described
earlier
is the feature of the paramagnetic state, which is devoid of vacancy order and should have an antiferromagnetic $J_1$.

Finally, higher-spin interactions such as a biquadratic term can also be incorporated in the model.
These interactions will generate
contributions to the free energy that are higher order in the pairing amplitudes. Therefore, they will not significantly affect the competition
among the different pairing channels.

\vspace{0.2cm}

\noindent{\bf Estimate of the superconducting energy gaps}

Our results allow an order-of-magnitude estimate of the pairing energy gaps
for the alkaline iron selenides and optimally doped iron pnictides.
For both the alkaline iron selenides and the iron pnictides, the energies of the zone-boundary
magnetic excitations are on the order of 200 meV,
which imply that the effective
exchange interactions are on the order of 20-50 meV \cite{MWang,Zhao}.
This specifies the order of magnitude of the parameters $\mathrm{J}_1$ and $\mathrm{J}_2$
in our model, even though they are for individual orbitals.
Our calculated pairing amplitudes for the pairing amplitudes $\Delta$ (Figs.~3c,d)
then imply that the
corresponding pairing gaps, $\sim 2 J \Delta$, appropriately weighted over the different orbitals,
will be on the order of ten meV. This is
compatible with what have been
inferred from the ARPES and tunneling measurements~\cite{YZhang,TQian,DMou,NATPHYS}.

\vspace{0.2cm}
\noindent
{\bf Pairing for the single-layer $\mathrm{\bf{Fe}}\mathrm{\bf{Se}}$ film.}
\label{slfs_pairing}

We consider the spin singlet pairing of a five-orbital
$t-J_1-J_2$ model for the newly discovered single-layer FeSe
film~\cite{FeSeSTO}. Recent experiments~\cite{FeSeSTO,DLiu12} suggest
that superconductivity arises from the FeSe layer, but not from the
FeSe/SrTiO$_3$ interface. Hence we study, using DFT, the electronic
structure of a single-layer FeSe without including a substrate.
We then use a five-orbital tight-binding
parametrization
described earlier in the Methods section
to fit the
DFT bandstructure. The best fitted tight-binding parameters are listed in
Supplementary Table~S3.
Supplementary
Figure~S6{\bf a}
shows the bandstructure of the tight-binding
model with these parameters. We have
fixed
the Fermi energy
so that the
 electron doping
 is 0.1 per Fe, which is close to the value of 0.09 per Fe
 determined by ARPES measurement~\cite{DLiu12}.
 As shown in Supplementary Figure~S6{\bf b},
 the calculated Fermi surface
 comprises electron pockets only, which is consistent with the
 ARPES results~\cite{DLiu12}. The overall bandwidth of this model is somewhat
 larger than
 that of
 \text{K$_{1-x}$Fe$_{2-y}$Se$_2$},
 suggesting the the electrons are
 less
 correlated in the single-layer FeSe than in alkaline iron selenides.
 This is consistent with the weaker mass renormalization
 observed in the single-layer FeSe film~\cite{DLiu12}.

Given these tight-binding parameters for the single-layer FeSe,
the corresponding five-orbital $t-J_1-J_2$ model
is specified as described earlier in the Methods section.
We again consider spin singlet pairing.
Supplementary Figure~S7
shows the pairing phase diagram and pairing amplitudes in several channels.
Both the phase diagram and the strength of the pairing amplitudes are similar to those of alkaline iron
selenides \KFS\ and \TFS\ .



\newpage


%

\vspace{0.5cm}

\noindent{\large\bf Acknowledgements}
The work was supported in part by
the NSF Grant No. DMR-1309531 and the Robert A. Welch Foundation
Grant No. C-1411 (R.Y., P.G. and Q.S.),
the National Science Foundation of China Grant No. 11374361 (R.Y.),
the Office of Naval Research Grant N00014-09-1-1025A and the National Institute of Standards and
Technology Grant 70NANB7H6138, Am001 (P.N.),
and the U.S. DOE under Contract No. DE-AC52-06NA25396,
the U.S. DOE Office of Basic Energy Sciences, and
the Center for Integrated Nanotechnologies ---a U.S. DOE user facility
(J.-X.Z.).

\vspace{0.5cm}

{\noindent\large\bf Author contributions}
All authors contributed substantially to this work.

\vspace{0.5cm}

{\noindent\large\bf Additional information}
The authors declare no competing financial interests.
Supplementary information accompanies this paper on
www.nature.com/naturecommunications.
Reprints and permissions information is available online at www.nature.com/reprints.
Correspondence and requests for
materials should be addressed to Q.S.  (qmsi@rice.edu).





\newpage
\begin{figure}[h!]
\caption{\label{fig1}
{\bf Schematic
phase diagram
near a Mott transition.}
In this zero-temperature phase diagram, the red point located on the $U/W$ axis refers to the
point of the Mott transition,
while the purple shading illustrates the regime that has strong antiferromagnetic correlations.
The parent compounds of alkaline iron selenides and iron pnictides
are located in the vicinity of, \emph{albeit} on the two sides of, the Mott transition.
Superconductivity occurs at nonzero carrier doping,
with the optimal doping located in the region indicated by the arrows.
}
\end{figure}

\begin{figure}[h!]
\caption{\label{fig2}
{\bf The contrasting Fermi surfaces of  \KFS\
and iron pnictides.}
{\bf a} and {\bf b} respectively show the Fermi surfaces of
\KFS\ and iron pnictides in the extended Brillouin zone corresponding to one iron per unit cell,
as obtained by using a five orbital tight binding model described in the
Methods section,
for electron doping $\delta=0.15$. There are only electron pockets at
the M
points $(\pm \pi,0)$ and $(0, \pm \pi)$ for
\text{K$_{1-x}$Fe$_{2-y}$Se$_2$}.
For iron pnictides, there are in addition two hole pockets near the $\mathrm{\Gamma}$ point $(0,0)$.
{\bf c} and {\bf d} show the corresponding orbital weights (O.W.)
on the electron
pockets centered at $(\pi,0)$ in {\bf a} and {\bf b}.
$\theta$ is the winding angle of the pocket with respect to its center.
The weights on the electron pocket centered
at $(0,\pi)$ can be obtained by interchanging the $xz$ and $yz$
components and shifting $\theta$ by $\pi/2$.
}
\end{figure}

\begin{figure}[h!]
\caption{\label{fig3}
{\bf  Phase
diagram and pairing amplitudes of
\KFS\ and iron pnictides.}
The results presented here are for electron doping $\delta=0.15$.
{\bf a} and {\bf b} respectively show the zero-temperature phase diagrams of
\KFS\ and iron pnictides. Region I corresponds to an $\mathrm{A}_{1g}$
state with dominant $s_{x^2y^2}$ channel. Regions II and III
mark an $\mathrm{A}_{1g}+i\mathrm{B}_{1g}$ state with
dominant $s_{x^2y^2}$ and $d_{x^2-y^2}$ channels (II) and
dominant $s_{x^2+y^2}$ and $d_{x^2-y^2}$ channels (III); the phase locking occurs
only at low temperatures.
Region IV for
\KFS\ is a pure $\mathrm{B}_{1g}$ state with dominant $d_{x^2-y^2}$ channel.
{\bf c} and {\bf d} display the corresponding pairing amplitudes for the $xy$
orbital of
\KFS\ and iron pnictides.
}
\end{figure}

\newpage

\begin{figure}[ht!]
\caption{\label{fig4}
{\bf Pairing amplitudes of
\KFS\ and iron pnictides for $xy$ and $xz/yz$ orbitals.}
{\bf a} and {\bf b}, comparison of the competing dominant pairing
amplitudes $\mathrm{A}_{1g}$ $s_{x^2y^2}$, $\mathrm{A}_{1g}$
$s_{x^2+y^2}$, and $\mathrm{B}_{1g}$ $d_{x^2-y^2}$ for the $xy$ orbital of
\KFS\ and iron pnictides, both for electron doping $\delta=0.15$ and $J_2/D=0.1$.
{\bf c} and {\bf d}, the same as {\bf a} and {\bf b} but for the $xz/yz$ orbitals. For
\KFS\,
the amplitude for the $\mathrm{A}_{1g}$ $s_{x^2+y^2}$ channel is strongly suppressed compared
to the pnictides case. Correspondingly, a pure $\mathrm{B}_{1g}$ $d_{x^2-y^2}$ state
is observed for
\KFS\ but is absent in iron pnictides.
}
\end{figure}


\newpage

\begin{figure}[H]
\centerline{\includegraphics[width=170mm]{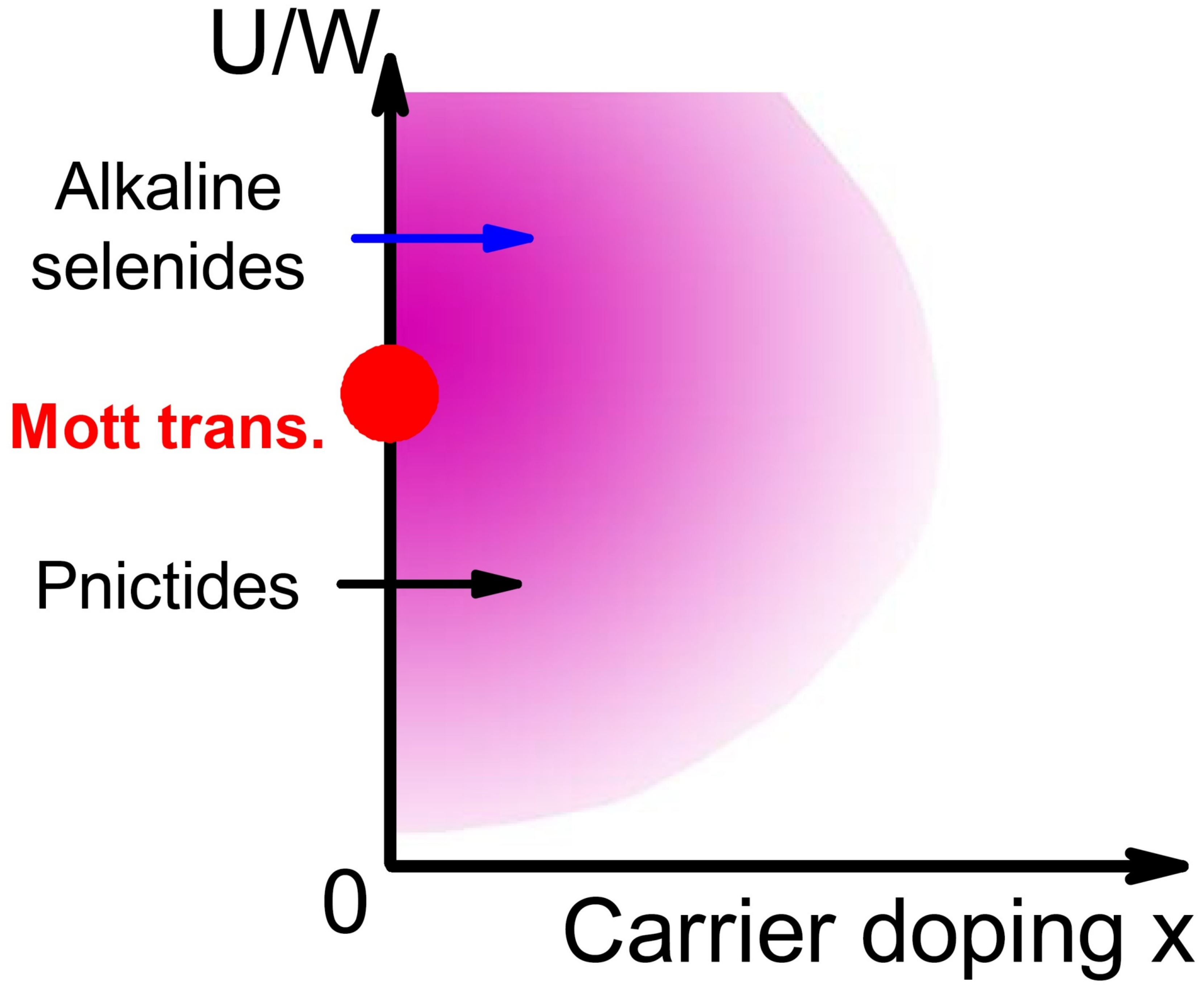}
}
{\Large Figure 1}
\vspace{8cm}
\end{figure}

\newpage

\begin{figure}[H]
\centerline{\includegraphics[width=170mm]{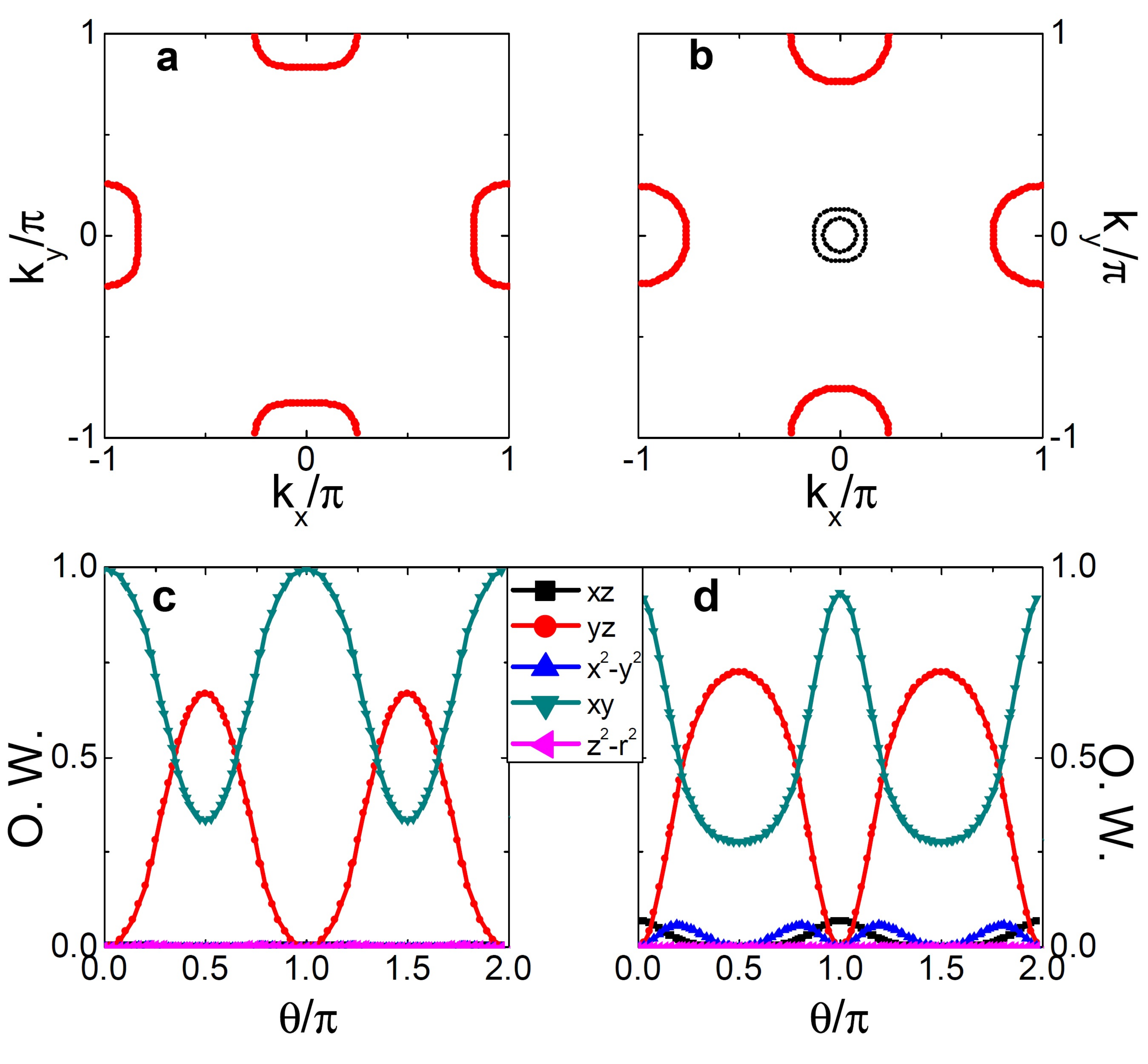}
}
{\Large Figure 2}
\vspace{8cm}
\end{figure}

\newpage

\begin{figure}[H]
\centerline{\includegraphics[width=183mm]{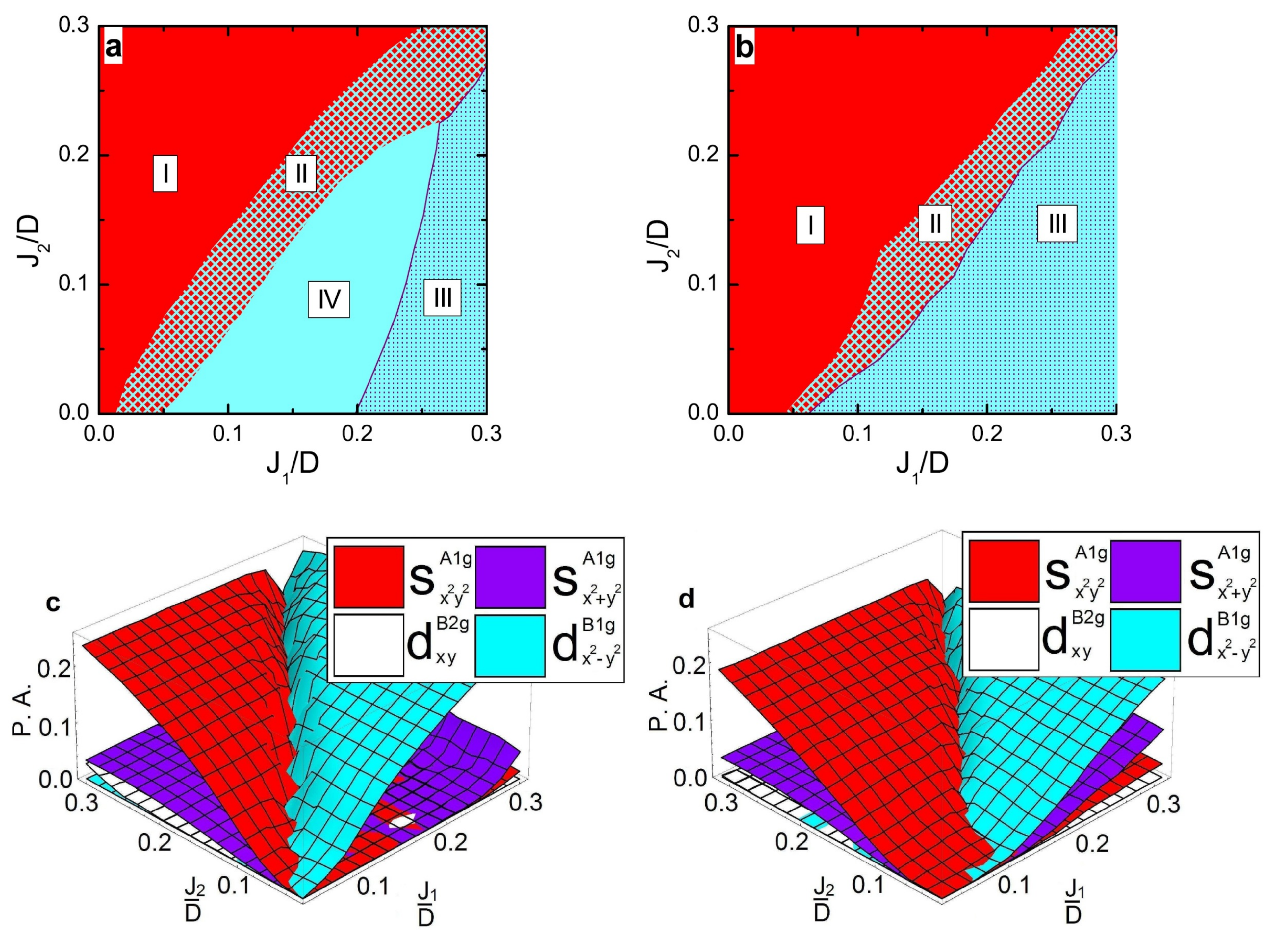}
}
{\Large Figure 3}
\vspace{8cm}
\end{figure}

\begin{figure}[H]
\centerline{\includegraphics[width=170mm]{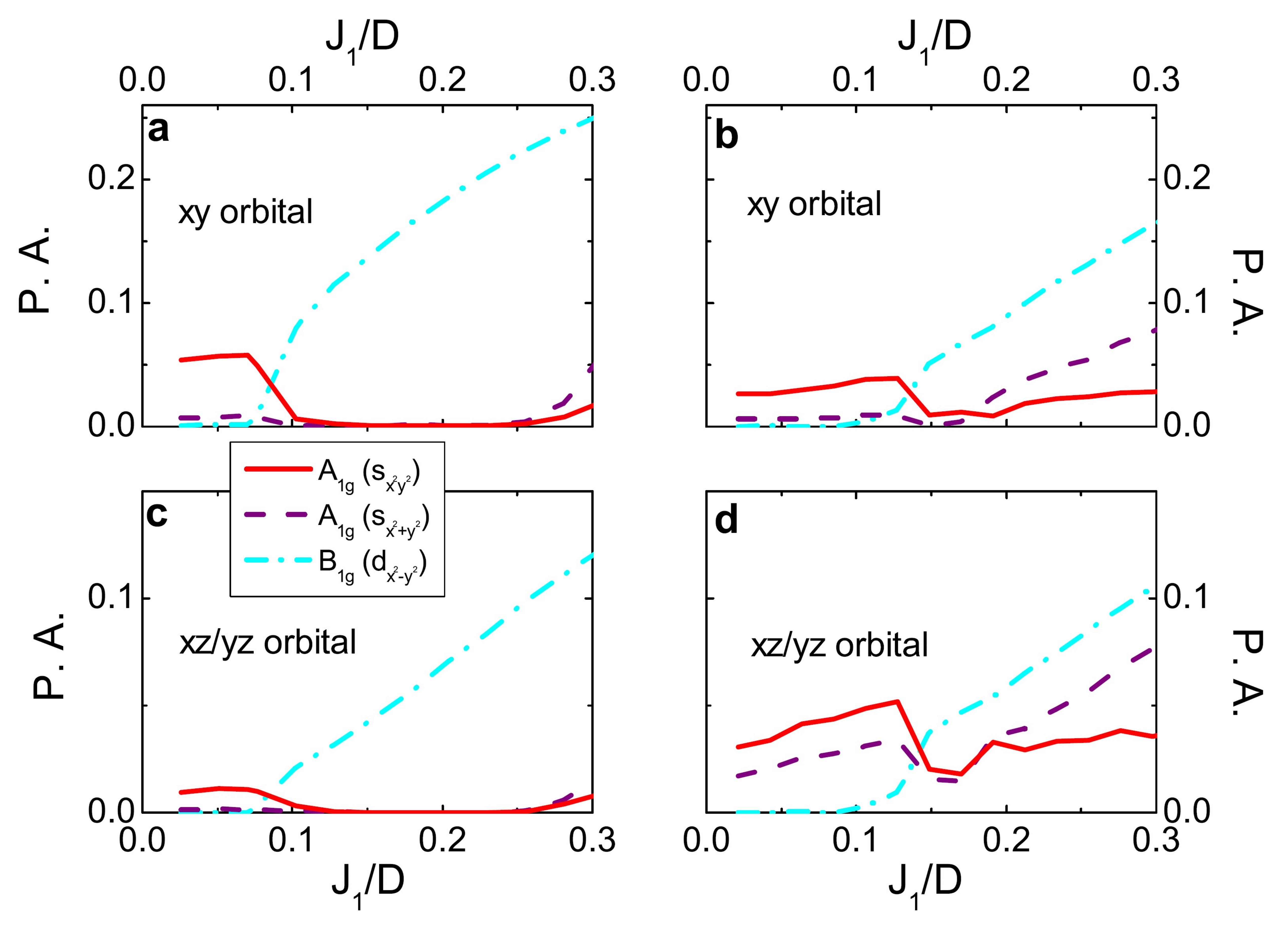}
}
{\Large Figure 4} 
\vspace{10cm}
\end{figure}


\def\Re{\text{Re}}
\def\Im{\text{Im}}
\def\sgn{\text{sgn}}
\renewcommand\figurename{{\bf Supplementary Figure S$\!\!$}}
\renewcommand\tablename{{\bf Supplementary Table S$\!\!$}}
\setcounter{figure}{0}


\onecolumngrid

\newpage
\noindent{\bf Supplementary Figures.}

\vskip 2 cm

\begin{figure}[H]
\centering
\includegraphics[
width=170mm
]{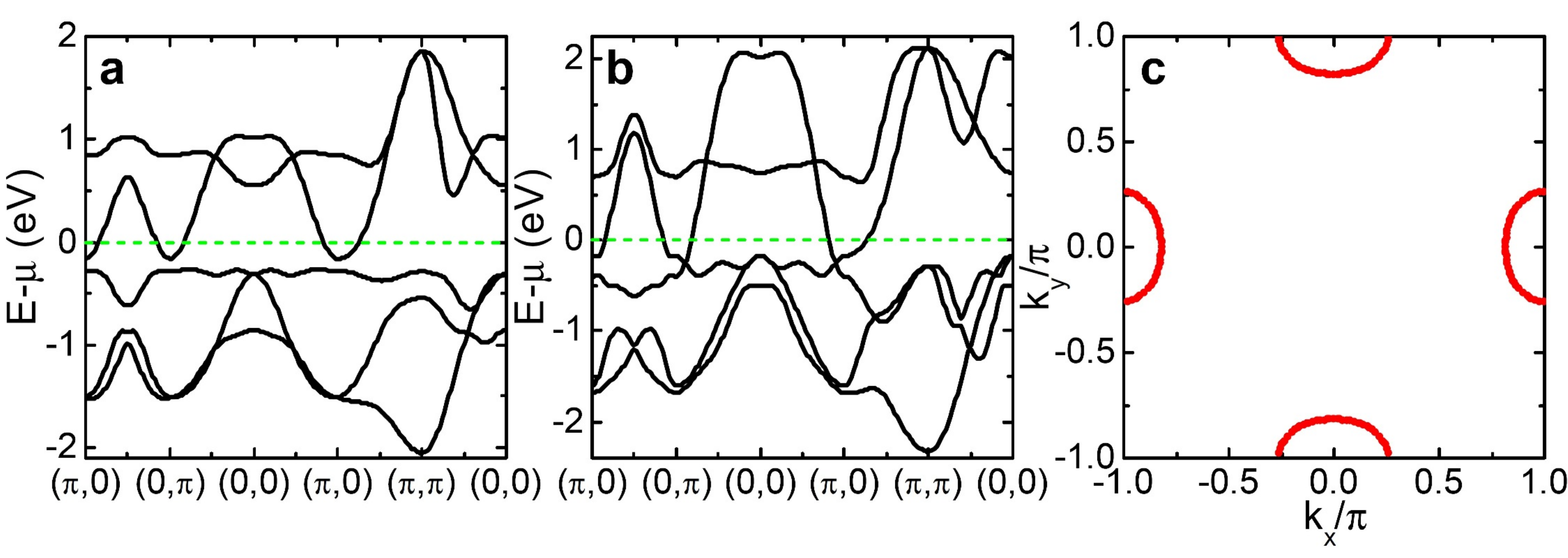}
\caption{The band dispersions of {\bf a} \KFS\ 
and {\bf b} \TFS, 
along the high-symmetry directions of the extended Brillouin zone (one iron/unit cell).
The band dispersions have been obtained by diagonalizing the appropriate five-orbital tight binding model.
Panel {\bf c} shows the Fermi surfaces of
\TFS\
with electron doping $\delta=0.15$, which consist of only electron pockets near zone boundaries M.
}\label{fig:S1}
\end{figure}

\newpage
\begin{figure}[H]
\centering
\includegraphics[
width=170mm
]{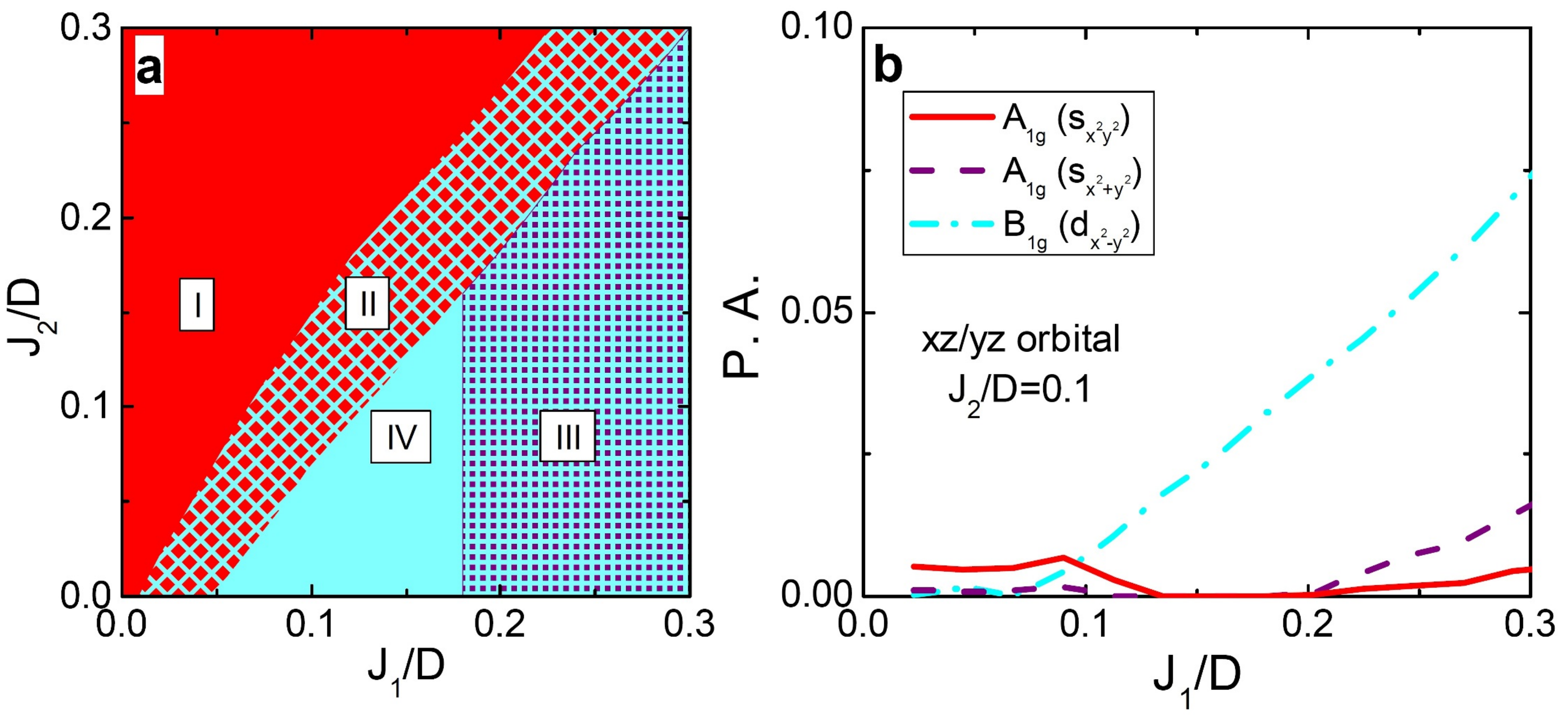}
\caption{{\bf Pairing phase diagram and amplitudes of
$\mathrm{\bf{Tl}}_{\bf{1-x}}\mathrm{\bf{Fe}}_{\bf{2-y}}\mathrm{\bf{Se}}_{\bf 2}$.}
The phase diagram and the strength of the pairing amplitudes are similar to those of
\KFS\,
described in the main text.
Panel {\bf a} shows the zero temperature phase diagram of
\TFS\
for an electron doping $\delta=0.15$.
The regions I, II, III, and IV respectively correspond
to an $\mathrm{A}_{1g}$ state
with $s_{x^2y^2}$ as the dominant pairing channel, a time reversal symmetry breaking
$\mathrm{A}_{1g}+i\mathrm{B}_{1g}$
state with $s_{x^2y^2}$ and $d_{x^2-y^2}$ as the dominant $\mathrm{A}_{1g}$ and $\mathrm{B}_{1g}$
pairing channels,
a likewise $\mathrm{A}_{1g}+i\mathrm{B}_{1g}$ state with $s_{x^2+y^2}$ and $d_{x^2-y^2}$
as the dominant $\mathrm{A}_{1g}$ and $\mathrm{B}_{1g}$ pairing channels,
and a pure $\mathrm{B}_{1g}$ state
with $d_{x^2-y^2}$ pairing channel. Panel {\bf b} shows the competing dominant pairing amplitudes
$\mathrm{A}_{1g}$ $s_{x^2y^2}$, $\mathrm{A}_{1g}$ $s_{x^2+y^2}$, $\mathrm{B}_{1g}$ $s_{x^2y^2}$
for ${xz}$/${yz}$ orbitals of
\TFS\
with an electron doping $\delta=0.15$, and $\mathrm{J_2/D}=0.1$.
}\label{fig:S2}
\end{figure}

\newpage
\begin{figure}[H]
\centering
\includegraphics[
width=170mm
]{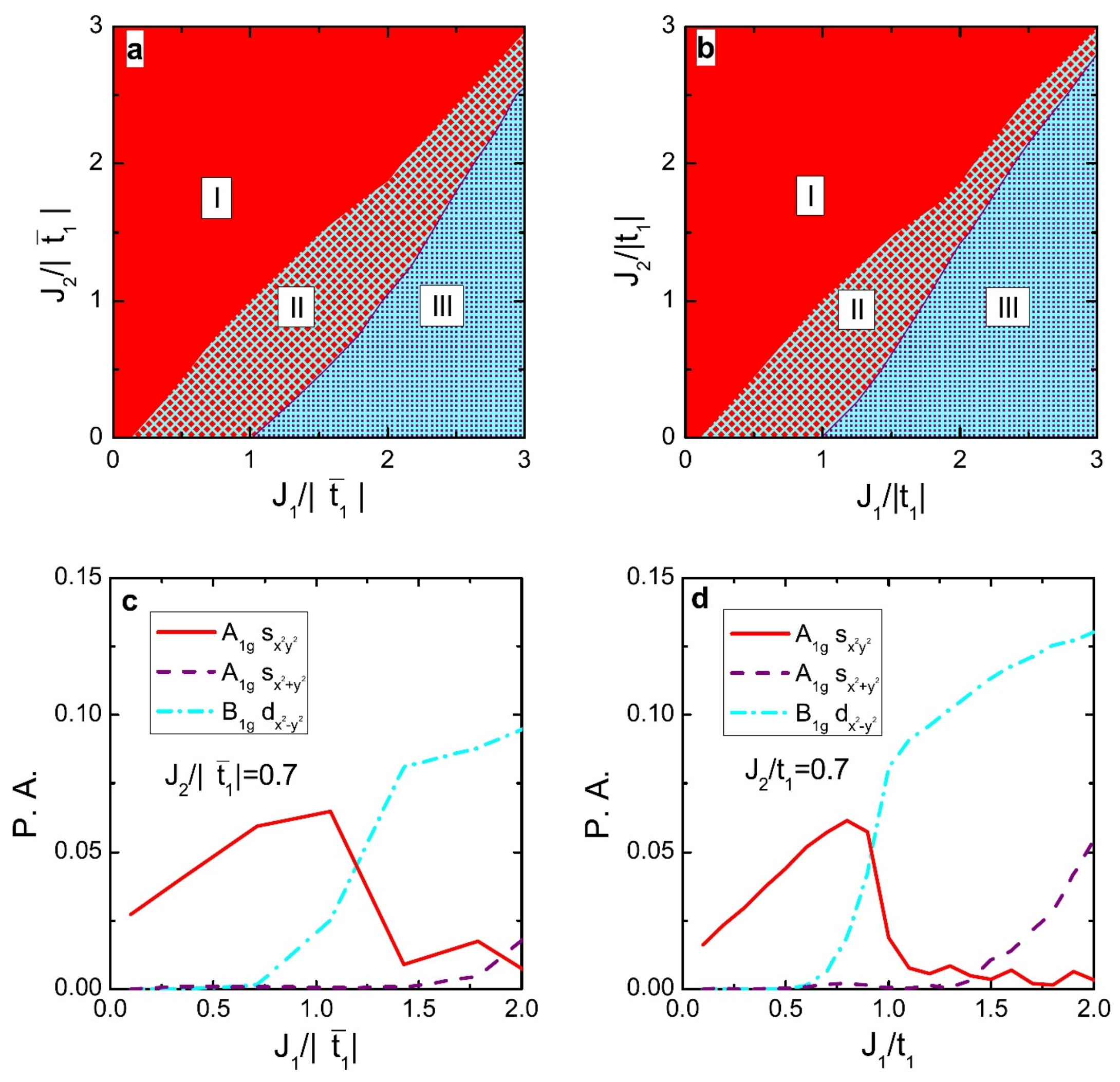}
\caption{Qualitatively similar zero temperature phase diagrams
of a two orbital $\mathrm{t}-\mathrm{J}_1-\mathrm{J}_2$
model obtained {\bf a} by considering a doping dependent renormalization
of the band dispersions due to the local constraint
of zero double occupancies and {\bf b}
without imposing the zero double occupancy constraint. In {\bf a} the effective
hopping parameters $\bar{t}_i=t_i\frac{\delta}{2}$. The regions I, II,
and III respectively correspond to an $\mathrm{A}_{1g}$
state with $s_{x^2y^2}$ as the dominant pairing channel,
an $\mathrm{A}_{1g}+i\mathrm{B}_{1g}$
state with $s_{x^2y^2}$ and $d_{x^2-y^2}$ as the dominant
$\mathrm{A}_{1g}$ and $\mathrm{B}_{1g}$
channels,
and a likewise  $\mathrm{A}_{1g}+i\mathrm{B}_{1g}$ state with $s_{x^2+y^2}$
and $d_{x^2-y^2}$
as the dominant $\mathrm{A}_{1g}$ and $\mathrm{B}_{1g}$
channels.
}\label{fig:S3}
\end{figure}

\newpage
\begin{figure}[H]
\centering
\includegraphics[
width=170mm
]{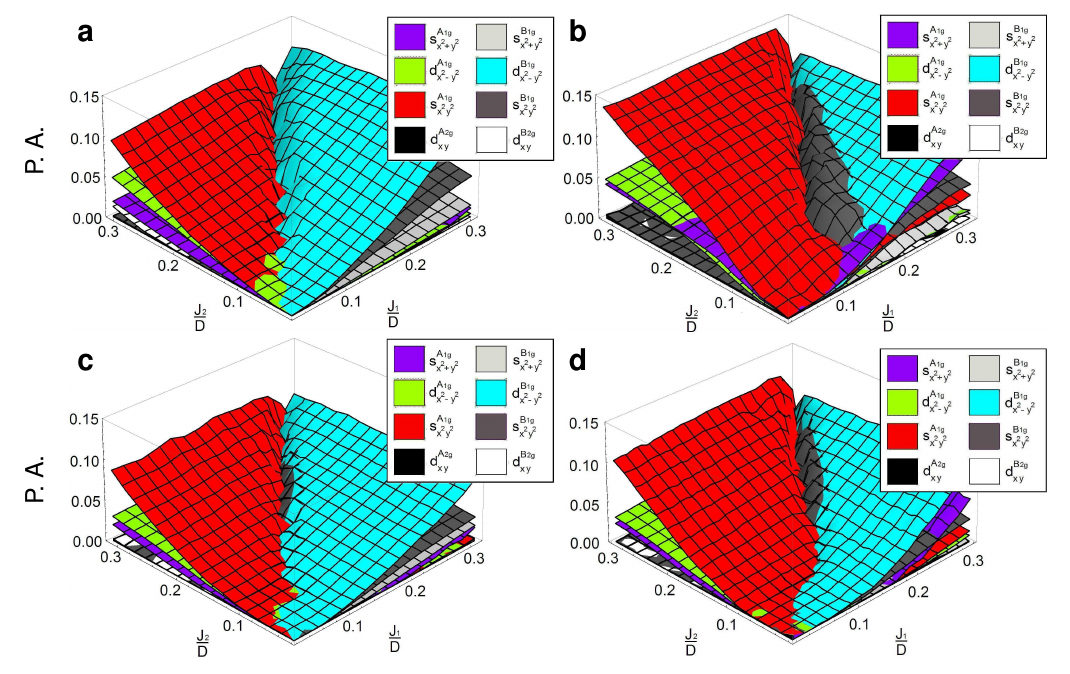}
\caption{The pairing amplitudes for $xz$/$yz$ orbitals of {\bf a} \KFS,
{\bf b} iron pnictides with electron doping $\delta=0.15$,
{\bf c} \TFS, and {\bf d} single-layer FeSe.
Compared to \KFS\ 
and \TFS, 
the iron pnictides have stronger $\mathrm{A}_{1g}$ $s_{x^2+y^2}$ pairing.
}\label{fig:S4}
\end{figure}

\newpage
\begin{figure}[H]
\centering
\includegraphics[
width=170mm
]{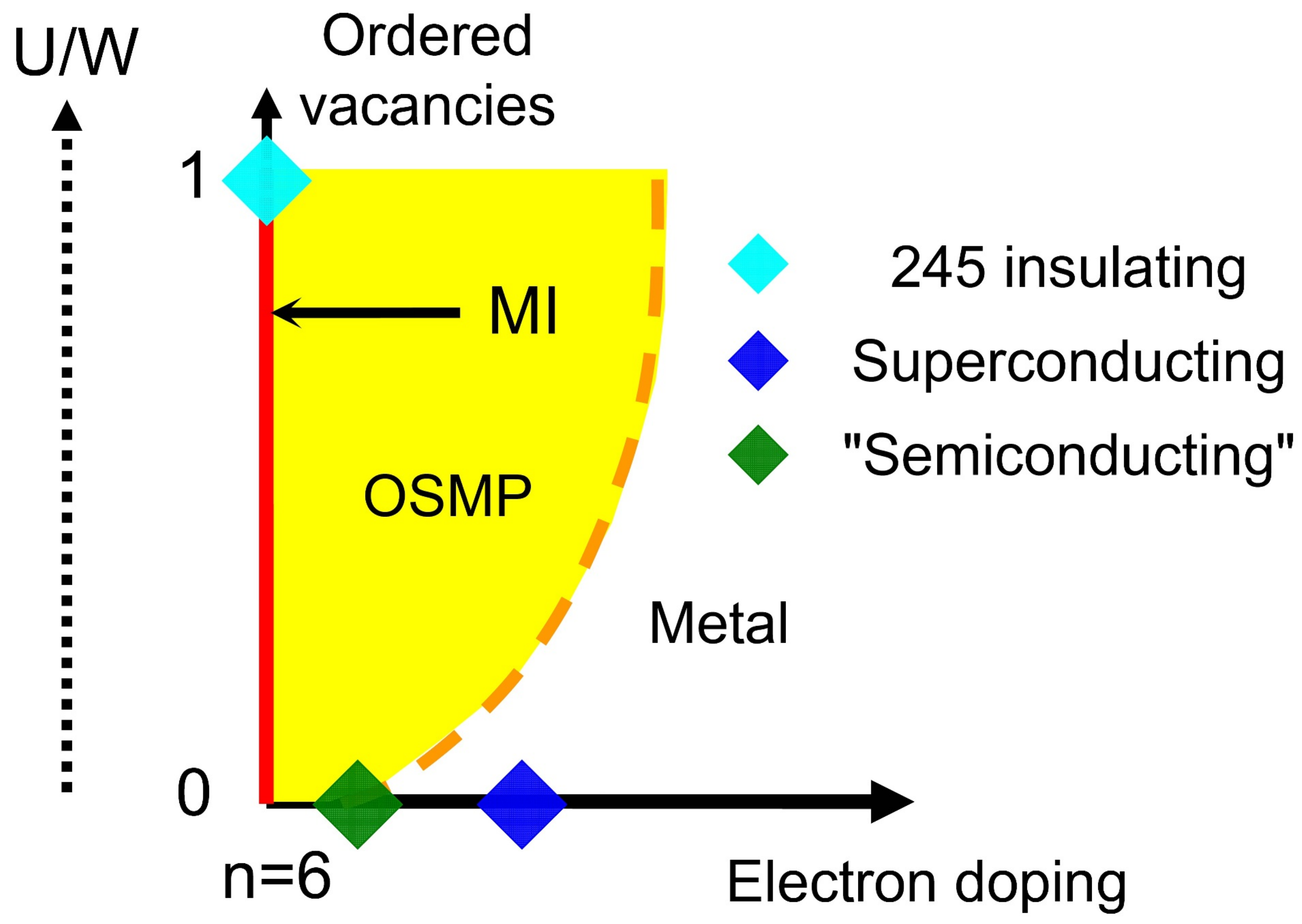}
\caption{Sketch of the overall phase diagram for
\text{K$_{1-x}$Fe$_{2-y}$Se$_2$}.
The vertical axis stands for the strength of iron vacancy order,
with 1 being fully $\sqrt{5}\times\sqrt{5}$ vacancy ordered and 0 being
completely vacancy disordered.
The iron vacancy order affects the system in a similar way as $U/W$.
The red line refers to the Mott insulator (MI),
the yellow shading illustrates the orbital-selective Mott phase (OSMP),
and the orange dashed line shows an OSMP-to-metal transition.
The diamond symbols indicate the approximate positions where the various samples are located.
}\label{fig:S5}
\end{figure}

\newpage
\begin{figure}[H]
\centering
\includegraphics[
width=170mm
]{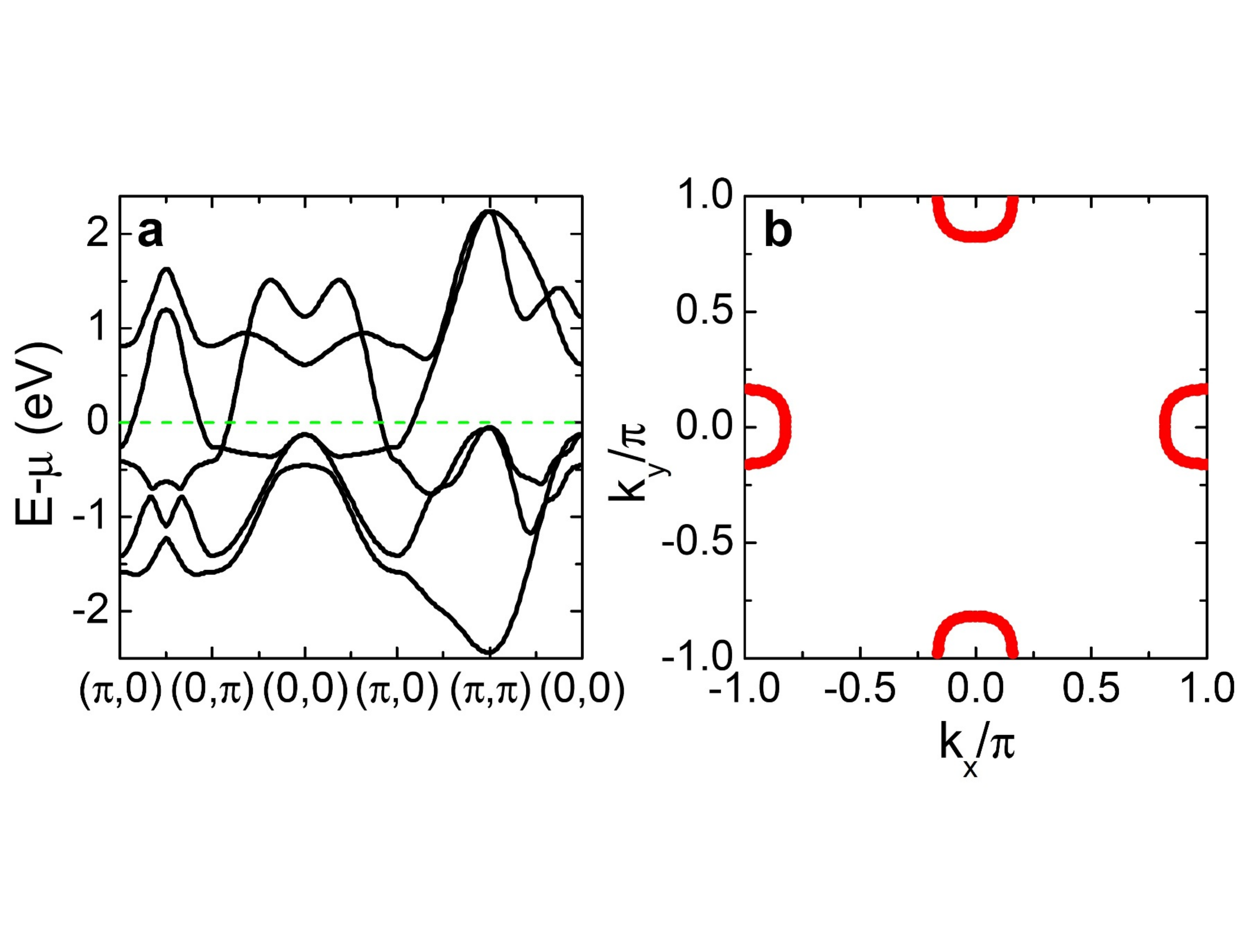}
\caption{Panel {\bf a} shows the band dispersion of the single-layer FeSe
along the high-symmetry directions of the extended Brillouin zone (one iron/unit cell).
The band dispersion has been obtained by diagonalizing the appropriate five-orbital tight binding model.
Panel {\bf b} shows the Fermi surfaces of
the single-layer FeSe
with electron doping $\delta=0.1$, which consist of only electron pockets near zone boundaries M.
}\label{fig:S6}
\end{figure}

\newpage
\begin{figure}[H]
\centering
\includegraphics[
width=170mm
]{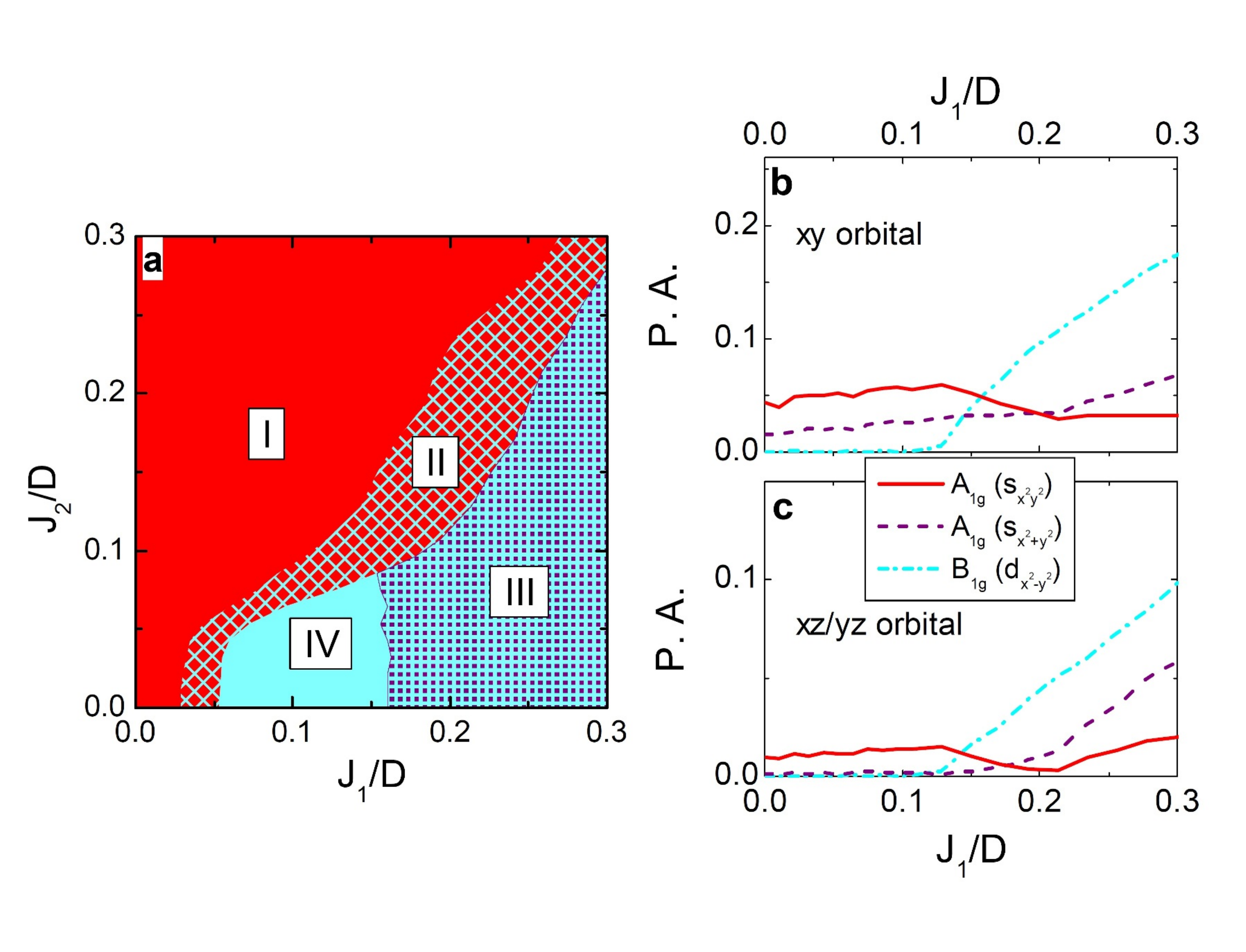}
\caption{Panel {\bf a} shows the zero temperature phase diagram of
the single-layer FeSe
for an electron doping $\delta=0.1$. The phase diagram is similar to those of
\KFS\
(Fig.~3{\bf a} of the main text) and \TFS\ (Supplementary Figure~S2{\bf a}).
The regions I, II, III, and IV respectively correspond
to an $\mathrm{A}_{1g}$ state
with $s_{x^2y^2}$ as the dominant pairing channel, a time reversal symmetry breaking
$\mathrm{A}_{1g}+i\mathrm{B}_{1g}$
state with $s_{x^2y^2}$ and $d_{x^2-y^2}$ as the dominant $\mathrm{A}_{1g}$ and $\mathrm{B}_{1g}$
pairing channels,
a likewise $\mathrm{A}_{1g}+i\mathrm{B}_{1g}$ state with $s_{x^2+y^2}$ and $d_{x^2-y^2}$
as the dominant $\mathrm{A}_{1g}$ and $\mathrm{B}_{1g}$ pairing channels,
and a pure $\mathrm{B}_{1g}$ state
with $d_{x^2-y^2}$ pairing channel. Panels {\bf b} and {\bf c} respectively
show the competing dominant pairing amplitudes
$\mathrm{A}_{1g}$ $s_{x^2y^2}$, $\mathrm{A}_{1g}$ $s_{x^2+y^2}$, $\mathrm{B}_{1g}$ $s_{x^2y^2}$
for $xy$ and ${xz}$/${yz}$ orbitals of
the single-layer FeSe
with electron doping $\delta=0.1$, and $\mathrm{J_2/D}=0.1$.
}\label{fig:S7}
\end{figure}

\newpage
\noindent{\bf Supplementary Tables.}

\vskip 2 cm

\begin{minipage}{\linewidth}
\begin{center}
\begin{tabular}{cccccccc}
  \hline
  \hline
    & $\alpha=1$ & $\alpha=2$ & $\alpha=3$ & $\alpha=4$ & $\alpha=5$ &   &   \\ \hline
  $\epsilon_\alpha$ & -0.36559 & -0.36559 & -0.56466 & -0.00096 & -0.91583 &  & \\ \hline\hline
  $t^{\alpha\alpha}_\mu$ & $\mu=x$ & $\mu=y$ & $\mu=xy$ & $\mu=xx$ & $\mu=xxy$ & $\mu=xyy$ & $\mu=xxyy$ \\ \hline
  $\alpha=1$ & -0.05475 & -0.40868 & --0.20881 & --0.01557 & -0.00866 & -0.03143 & --0.01899\\ \hline
  $\alpha=3$ & --0.32523 &  & -0.09783 & -0.00537 &  &  &  \\ \hline
  $\alpha=4$ & --0.20633 &  & --0.06582 & -0.03525 & -0.02189 &  & --0.00423 \\ \hline
  $\alpha=5$ & -0.04270 &  &  & --0.01117 & --0.00177 &  & -0.01349 \\ \hline\hline
  $t^{\alpha\beta}_\mu$ & $\mu=x$ & $\mu=xy$ & $\mu=xxy$ & $\mu=xxyy$ &  &  &  \\ \hline
  $\alpha\beta=12$ &  & --0.10161 & -0.02017 & --0.03273 &  &  &  \\ \hline
  $\alpha\beta=13$ & -0.31447 & --0.06225 & --0.01030 &  &  &  &  \\ \hline
  $\alpha\beta=14$ & --0.13785 & --0.00105 & --0.01040 &  &  &  &  \\ \hline
  $\alpha\beta=15$ & -0.04825 & -0.10096 &  & -0.01204 &  &  &  \\ \hline
  $\alpha\beta=34$ &  &  & -0.04795 &  &  &  &  \\ \hline
  $\alpha\beta=35$ & -0.30966 &  & -0.01498 &  &  &  &  \\ \hline
  $\alpha\beta=45$ &  & -0.08359 &  & -0.00766 &  &  &  \\ \hline
  \hline
\end{tabular}
\par
\end{center}
\bigskip
\noindent
{\bf Supplementary Table S}1.
 Tight-binding parameters of the five-orbital model for
\text{K$_{1-x}$Fe$_{2-y}$Se$_2$}.
Here we
use the same notation as in Ref.~
27 of the main text.
The orbital index $\alpha=$1,2,3,4,5 correspond to $d_{xz}$, $d_{yz}$,
$d_{x^2-y^2}$, $d_{xy}$, and $d_{3z^2-r^2}$ orbitals, respectively.
The units of the parameters are eV.
\end{minipage}

\newpage
\begin{minipage}{\linewidth}
\begin{center}
\begin{tabular}{cccccccc}
  \hline
  \hline
    & $\alpha=1$ & $\alpha=2$ & $\alpha=3$ & $\alpha=4$ & $\alpha=5$ &   &   \\ \hline
  $\epsilon_\alpha$ & -0.35956 & -0.35956 & -1.11574 & 0.09324 & -0.74545 &  & \\ \hline\hline
  $t^{\alpha\alpha}_\mu$ & $\mu=x$ & $\mu=y$ & $\mu=xy$ & $\mu=xx$ & $\mu=xxy$ & $\mu=xyy$
  & $\mu=xxyy$ \\ \hline
  $\alpha=1$ & -0.24198 & -0.34713 & 0.23289 & 0.10214 & -0.05889 & 0.06785 & 0.00728\\ \hline
  $\alpha=3$ & 0.38401 &  & -0.02547 & 0.01306 &  &  &  \\ \hline
  $\alpha=4$ & 0.38169 &  & 0.15837 & -0.03082 & -0.04663 &  & -0.02685 \\ \hline
  $\alpha=5$ & 0.03688 &  &  & -0.02562 & -0.03218 &  & 0.02102 \\ \hline\hline
  $t^{\alpha\beta}_\mu$ & $\mu=x$ & $\mu=xy$ & $\mu=xxy$ & $\mu=xxyy$ &  &  &  \\ \hline
  $\alpha\beta=12$ &  & 0.24408 & 0.00436 & 0.01173 &  &  &  \\ \hline
  $\alpha\beta=13$ & -0.36704 & -0.02212 & -0.02627 &  &  &  &  \\ \hline
  $\alpha\beta=14$ & 0.23177 & -0.05602 & 0.03873 &  &  &  &  \\ \hline
  $\alpha\beta=15$ & -0.13172 & -0.06082 &  & 0.02931 &  &  &  \\ \hline
  $\alpha\beta=34$ &  &  & -0.05093 &  &  &  &  \\ \hline
  $\alpha\beta=35$ & -0.23085 &  & 0.03286 &  &  &  &  \\ \hline
  $\alpha\beta=45$ &  & -0.17070 &  & -0.02720 &  &  &  \\ \hline
  \hline
\end{tabular}
\par
\end{center}
\bigskip
\noindent
{\bf Supplementary Table S}2.
Tight-binding parameters of the five-orbital model for
\text{Tl$_{1-x}$Fe$_{2-y}$Se$_2$}.
\end{minipage}

\newpage
\begin{minipage}{\linewidth}
\begin{center}
\begin{tabular}{cccccccc}
  \hline
  \hline
    & $\alpha=1$ & $\alpha=2$ & $\alpha=3$ & $\alpha=4$ & $\alpha=5$ &   &   \\ \hline
  $\epsilon_\alpha$ & -0.03123 & -0.03123 & -0.51304 & 0.43279 & -0.374 &  & \\ \hline\hline
  $t^{\alpha\alpha}_\mu$ & $\mu=x$ & $\mu=y$ & $\mu=xy$ & $\mu=xx$ & $\mu=xxy$ & $\mu=xyy$ & $\mu=xxyy$ \\ \hline
  $\alpha=1$ & -0.15497 & -0.34438 & 0.23647 & 0.0363 & -0.04761 & 0.00148 & 0.03479\\ \hline
  $\alpha=3$ & 0.38181 &  & -0.04947 & -0.05064 &  &  &  \\ \hline
  $\alpha=4$ & 0.22855 &  & 0.11723 & -0.02079 & -0.0402 &  & -0.0729 \\ \hline
  $\alpha=5$ & 0.01989 &  &  & 0.00197 & -0.03545 &  & 0.02824 \\ \hline\hline
  $t^{\alpha\beta}_\mu$ & $\mu=x$ & $\mu=xy$ & $\mu=xxy$ & $\mu=xxyy$ &  &  &  \\ \hline
  $\alpha\beta=12$ &  & 0.14939 & -0.00943 & 0.03927 &  &  &  \\ \hline
  $\alpha\beta=13$ & -0.35208 & 0.09872 & 0.06043 &  &  &  &  \\ \hline
  $\alpha\beta=14$ & 0.22397 & -0.08589 & 0.02547 &  &  &  &  \\ \hline
  $\alpha\beta=15$ & -0.08056 & -0.06893 &  & -0.01788 &  &  &  \\ \hline
  $\alpha\beta=34$ &  &  & -0.00317 &  &  &  &  \\ \hline
  $\alpha\beta=35$ & -0.36967 &  & -0.03581 &  &  &  &  \\ \hline
  $\alpha\beta=45$ &  & -0.23453 &  & 0.02949 &  &  &  \\ \hline
  \hline
\end{tabular}
\end{center}
\bigskip
\noindent
{\bf Supplementary Table S}3.
Tight-binding parameters of the five-orbital model for the single-layer \text{FeSe}.
\end{minipage}
%

\newpage
\noindent{\bf Supplementary Note 1.
Two orbital model with and without zero double-occupancy constraints}
\label{constraint}

To demonstrate the robustness of our pairing phase diagram against
the doping dependent band renormalization effects,
we consider a two orbital model involving only $xz$ and $yz$ orbitals. For simplicity
we consider hole doping;
the electron doping case can be treated in a similar manner after performing a particle-hole
transformation. The model and the detail of the theoretical method
are described in the Supplementary Methods.

We have compared both the pairing phase diagrams and pairing amplitudes
obtained using
the two methods described in the Supplementary Methods.
One method explicitly imposes the no double-occupancy constraint, and the other accounts for it through
an effective band renormalization.
As shown in Supplementary Figure~S\ref{fig:S3}, the pairing phase diagrams and the pairing amplitudes
obtained
in these two methods
are qualitatively similar, when the exchange interactions are scaled by the renormalized bandwidth $D$.

\vskip 0.2 cm

\noindent{\bf Supplementary Note 2. Orbital character of the Fermi surface and orbital dependence of pairing amplitudes.}

We have respectively shown in Fig.~2c and Fig.~2d of the main text the orbital weights
on the electron pockets for both
\KFS\
and pnictides, for
electron doping $\delta=0.15$. For pnictides, the $xz$/$yz$ orbitals have the dominant  orbital weights
on the two hole pockets (not shown), and also contribute significantly on the electron pockets.
Whereas for
\KFS\
(and also for
\TFS\ ), there are no hole pockets, and the contribution from $xy$ orbital is considerably enhanced on the electron pockets.
This indicates that the $xy$ orbital plays a more important role in building the electron pocket
of 
\KTFS\ than in pnictides.

The enhanced $xy$ orbital character on Fermi surface of
\KFS\ affects the electron pairing. To see this we compare the pairing amplitudes at $J_2/D=0.1$
of $xy$ orbital with $xz$/$yz$ orbital for both
\KFS\
and pnictides at electron doping $\delta=0.15$ in Fig.~4 of the main text.
We see that for pnictides, the contribution to the three competing dominant pairings from the $xz$/$yz$
orbital is comparable with the one from $xy$ orbital. But for
\KFS\, the contribution from $xy$ orbital can be stronger. This is especially true in the regime
$J_1/D\lesssim 0.07$, where only $\mathrm{A}_{1g}$ pairing is present.

Overall, however, the pairing amplitudes for the $xy$ orbital are similar between
\linebreak
\KTFS\
and iron pnictides. This is seen in the three-dimensional plots given in Figs.~3c,d of the main text and
in the Supplementary
 Figure~S\ref{fig:S2}b, which are similar to their counterparts for the $xz$/$yz$ orbitals (Supplementary
 Figure~S\ref{fig:S4}).

\vskip 0.2 cm

\noindent{\bf Supplementary Note 3. Mott phase with vacancy order
and the orbital selective Mott phase in alkaline iron selenides.}
\label{osmt}

Experiments reveal that the insulating alkaline iron selenide samples contain
$\sqrt{5}\times\sqrt{5}$
ordered iron vacancies. Many measurements also suggest that the superconducting
region of the superconducting samples
include
either no iron vacancy, or vacancies that are disordered. This raises the question of what role the vacancy ordered
insulating
phase (in the so called 245 compound) plays in the superconductivity
of the alkaline iron selenides,
and
how it can be connected to the superconducting phase.

In this note we
describe this linkage
within a phase diagram (Supplementary
 Figure~S\ref{fig:S5})
 in the parameter space spanned by $U/W$ associated
with
a varying degree of vacancy order and carrier doping,
derived from the considerations of
electron correlations in a multi-orbital model for
the alkline iron selenides  (Ref.~33 of the main text).
This phase diagram is supported by ARPES measurements (Ref.~34 of the main text),
 as well as transport measurements at high pressures  (Ref.~35 of the main text).
The horizontal and vertical axes respectively denote the electron doping and the strength of the iron
vacancy order.
We label the degree of vacancy order from 0 (being completely vacancy disordered) to 1
(for the fully vacancy ordered case).
Since the vacancy order causes the effective reduction of kinetic energies,
from the perspective of electron correlations,
it affects the system in a similar way as tuning the ratio $U/W$.
At commensurate filling $n=6$, we find the system
is in a Mott insulator (MI). Away from this filling, the system is in either an orbital-selective Mott phase
(OSMP) or a metallic state,
depending on the degree of vacancy order and the electron filling. In the OSMP, the 3d$_{xy}$
orbital is Mott localized,
while other 3d orbitals are still delocalized. In this phase diagram, the insulating 245 compounds
is located in the MI with full vacancy order.

A recent ARPES study (Ref.~34 of the main text)
suggests that the normal state of the superconducting sample and another sample
which shows semiconducting behavior in its resistivity are respectively located inside the metallic
phase and very close to the boundary of the
OSMP-to-metal transition, both without vacancy order but at different doping concentrations.
This suggests the following effects of chemical doping the 245 insulating sample: On the one hand, doping injects extra
carriers which increase the electron filling; on the other hand, the dopants change the chemical
environments around the iron ions,
and disturb the iron vacancy order. As a result of the combined effects, the system may evolve
from the vacancy ordered 245 insulating
phase to the vacancy disordered metallic one via the partially vacancy ordered OSMP. In this way,
the vacancy ordered 245 Mott insulating phase
connects to
the
superconducting material,
with the link provided by the OSMP (Ref.~35 of the main text).

\vskip 0.2 cm

\noindent
{\bf Supplementary Methods}\\
Here we describe the method used to study the singlet pairing in the two-orbital model.
The Hamiltonian of interest is
\begin{equation*}
H=-\sum_{i<j,\alpha,\beta,\sigma} t_{ij}^{\alpha \beta}c^{\dagger}_{i\alpha \sigma}
c_{j\beta \sigma}+h.c.
-\mu \sum_{i,\alpha,\sigma}c^{\dagger}_{i\alpha \sigma}c_{i\alpha \sigma}
+\sum_{i<j,\alpha,\beta} J_{ij}^{\alpha \beta} \left(\vec{S}_{i\alpha}\cdot
\vec{S}_{j\beta}-\frac{1}{4}n_{i\alpha} n_{j\beta}\right)
\tag{S1}
\end{equation*}
with the double occupancy prohibiting constraint
$\sum_{\sigma}c^{\dagger}_{i\alpha \sigma}c_{i\alpha \sigma}\leq1$ for each orbital.
The constraint is imposed by introducing slave boson operator $b_i$ and fermionic
spinon operator $f_i$
for each orbital (Refs.~21 and 45)
and the Hamiltonian is transformed to
\begin{align*}
H &=& -\sum_{i<j,\alpha,\beta,\sigma} t_{ij}^{\alpha \beta}f^{\dagger}_{i\alpha \sigma}
b_{i\alpha}b_{j\beta}^{\dagger}f_{j\beta \sigma}+h.c. -\mu \sum_{i,\alpha,\sigma}
f^{\dagger}_{i\alpha \sigma}f_{i\beta \sigma}-\sum_{i<j,\alpha,\beta} \frac{J_{ij}^{\alpha \beta}}{2}
B_{ij,\alpha \beta}^{\dagger}B_{ij,\alpha \beta} \nonumber\\
& & +\sum_{i,\alpha} \lambda_{i\alpha}\left(\sum_{\sigma}f^{\dagger}_{i\alpha \sigma}
f_{i\alpha \sigma}+b_{i\alpha}^{\dagger}b_{i\alpha}-1\right)
\tag{S2}
\end{align*}
where $B_{ij,\alpha \beta}^{\dagger}=(f^{\dagger}_{i\alpha \uparrow}f^{\dagger}_{j\beta \downarrow}
-f^{\dagger}_{i\alpha \downarrow}f^{\dagger}_{j\beta\uparrow})$
is the spin-singlet pairing operator for the fermionic spinons, and $\lambda_{i,\alpha}$'s
are Lagrange multipliers which
enforce the occupancy constraints. At zero temperature the slave bosons are Bose condensed
and the boson operators have finite expectation values.
In the tetragonal symmetry breaking particle-hole channel order, we have $\langle b_{i,1}\rangle
= \langle b_{i,2}\rangle=\sqrt{\delta/2}$, where $\delta$
is the hole doping and $\langle \lambda_{i,1}\rangle=\langle \lambda_{i,2}\rangle=\lambda$,
which is absorbed into the chemical potential.
The expectation value of the boson operators renormalizes the kinetic energy term by a factor of $\delta/2$.
As in the main text we choose
the interactions to be diagonal in the orbital space. After mean-field decoupling, the free energy density
\begin{align*}
f&=&\frac{J_1}{2}\sum_{\alpha}(|\Delta_{x,\alpha \alpha}|^2+|\Delta_{y,\alpha \alpha}|^2)+\frac{J_2}{2}
\sum_{\alpha}(|\Delta_{x+y, \alpha \alpha}|^2+|\Delta_{x-y,\alpha \alpha}|^2)\nonumber\\
&&-\int\frac{d^2k}{4\pi^2}(\mathcal{E}_{\mathbf{k}+}+\mathcal{E}_{\mathbf{k}-}-\frac{\delta}{2}
E_{\mathbf{k}+}-\frac{\delta}{2}E_{\mathbf{k}-}+2\mu)
\tag{S3}
\end{align*}
is minimized with respect to pairing amplitudes with the constraint $n_1=n_2=1-\frac{\delta}{2}$.
The quasiparticle dispersions $\mathcal{E}_{\mathbf{k}\pm}$ in the paired state are
calculated from the $4\times4$ Nambu matrix
\begin{equation*}
\hat{h}_{\mathbf{k}}=\left[\begin{array}{cc}\frac{\delta\hat{\xi}_{\mathbf{k}}}{2}-\mu \mathbbm{1}_{2\times2}
& \tilde{\mathbf{\Delta}}_{\mathbf{k}}\\\tilde{\mathbf{\Delta}}^{\ast}_{\mathbf{k}}
& -\frac{\delta \hat{\xi}_{\mathbf{k}}}{2}+\mu \mathbbm{1}_{2\times2}\end{array}\right]
\tag{S4}
\end{equation*}
where
\begin{align*}
\tilde{\Delta}_{\mathbf{k},\alpha \alpha}&=&J_1\left(\Delta_{x,\alpha \alpha}\cos(k_x)
+\Delta_{y,\alpha \alpha}\cos(k_y)\right)+J_2\left(\Delta_{x+y,\alpha \alpha}\cos(k_x+k_y)
+\Delta_{x-y,\alpha \alpha}\cos(k_x-k_y)\right)\nonumber\\
\tag{S5}
\end{align*}
For the band structure we choose the two orbital tight-binding model of Ref. \onlinecite{Raghu}.
The $2\times2$ tight-binding matrix $\hat{\xi}_{\mathbf{k}}=\hat{\xi}_{\mathbf{k}}
=\xi_{{\mathbf{k}}+}\mathbbm{1}_{2\times2}+\xi_{{\mathbf{k}}-}\tau_{z}
+\xi_{{\mathbf{k}}xy}\tau_{x}$, where Pauli matrices $\tau_i$)
operate on the orbital indices, and $\xi_{{\mathbf{k}}+}
=-(t_1+t_2)(\cos k_x+\cos k_y)-4t_3\cos k_x \cos k_y$,
$\xi_{{\mathbf{k}}-}=-(t_1-t_2)(\cos k_x-\cos k_y)$,
$\xi_{{\mathbf{k}}xy}=-4t_4\sin k_x \sin k_y$ are respectively $\mathrm{A}_{1g}$, $\mathrm{B}_{1g}$
and $\mathrm{B}_{2g}$ functions. The band dispersion relations $E_{{\mathbf{k}}\pm}
=\xi_{{\mathbf{k}}+}
\pm \sqrt{\xi_{{\mathbf{k}}-}^2+\xi_{{\mathbf{k}}xy}^2}$, give rise to two electron pockets
at ${\mathbf{k}}=(\pi,0)$ and $(0,\pi)$, and two hole pockets
at ${\mathbf{k}}=(0,0)$ and
$(\pi,\pi)$. The following values of the hopping parameters, $t_1=-t$, $t_2=1.3t$, $t_3=t_4=-0.85t$
were obtained in Ref.~\onlinecite{Raghu}, by a fitting of the LDA bands.
The quasiparticle dispersions in the paired state are given by
\begin{align*}
\mathcal{E}_{\mathbf{k}\pm}&=&\Bigg[\left(\left(\frac{\delta \xi_{\mathbf{k}+}}{2}-\mu \right)^{2}
+\frac{\delta^2}{4}\left(\xi_{\mathbf{k}-}^{2}+\xi_{\mathbf{k}xy}^{2}\right)
+\frac{|\tilde{\Delta}_{\mathbf{k},11}|^2}{2}
+\frac{|\tilde{\Delta}_{\mathbf{k},22}|^2}{2}\right)\pm \Bigg\{\bigg(\delta \xi_{\mathbf{k}-}
\left(\frac{\delta \xi_{\mathbf{k}+}}{2}-\mu \right)\nonumber \\&&+\frac{|\tilde{\Delta}_{\mathbf{k},11}|^2}{2}
-\frac{|\tilde{\Delta}_{\mathbf{k},22}|^2}{2}\bigg)^2+\delta^2\xi_{\mathbf{k}xy}^2
\left(\left(\frac{\delta \xi_{\mathbf{k}+}}{2}
-\mu \right)^2+\frac{1}{4}|\tilde{\Delta}_{\mathbf{k},11}-\tilde{\Delta}_{\mathbf{k},22}|^2\right)
\Bigg\}^{\frac{1}{2}}\Bigg]^{\frac{1}{2}}
\tag{S6}
\end{align*}

We then compare the pairing phase diagrams obtained from two means: one way is to explicitly account for the occupancy constraints and associated band renormalization; and the other way is to calculate the pairing
without imposing the no double-occupancy constraint.

\noindent{\bf Supplementary References}

\end{document}